\documentclass[12pt]{article}
\usepackage{graphics}
\usepackage{color}
\usepackage{amssymb,amsmath}
\usepackage{epsfig}
\newcommand{\bea}{\begin{eqnarray}}
\newcommand{\eea}{\end{eqnarray}}

 \makeatletter
\def\alt{\mathrel{\mathpalette\gl@align<}}
\def\agt{\mathrel{\mathpalette\gl@align>}}
\def\gl@align#1#2{\lower.6ex\vbox{\baselineskip\z@skip\lineskip\z@
\ialign{$\m@th#1\hfil##\hfil$\crcr#2\crcr\sim\crcr}}} \makeatother

\begin{document}
\begin{center}
\baselineskip 20pt 
{\Large\bf 
Inverse Seesaw Neutrino Signatures at LHC and ILC
}
\vspace{1cm}

{\large 
Arindam Das\footnote{adas8@ua.edu} 
and 
Nobuchika Okada\footnote{okadan@ua.edu}
} \vspace{.5cm}

{\baselineskip 20pt \it
Department of Physics and Astronomy, 
University of Alabama, \\ 
Tuscaloosa,  Alabama 35487, USA 
}

\vspace{.5cm}

\vspace{1.5cm} {\bf Abstract}
\end{center}
%%%%%%%%%%%%%%%%%%%%%%%%%%%%%%%%%%%
We study the collider signature of pseudo-Dirac heavy neutrinos 
 in the inverse seesaw scenario, 
 where the heavy neutrinos with mass at the electroweak scale 
 can have sizable mixings with the Standard Model neutrinos, 
 while providing the tiny light neutrino masses 
 by the inverse seesaw mechanism. Based on a simple, concrete model realizing the inverse seesaw, 
 we fix the model parameters 
 so as to reproduce the neutrino oscillation data and 
 to satisfy other experimental constraints, 
 assuming two typical flavor structures of the model 
 and the different types of hierarchical light 
 neutrino mass spectra. For completeness, we also consider a general parameterization 
 for the model parameters by introducing an arbitrary 
 orthogonal matrix, and the non-zero Dirac and Majorana phases.
We perform parameter scan to identify an allowed parameter region 
 which satisfies all experimental constraints. 
With the fixed parameters in this way, we analyze 
 the heavy neutrino signal at the LHC 
 through tri-lepton final states with large missing energy 
 and at the ILC through a single lepton plus di-jet 
 with large missing energy. We find that in some cases, the heavy neutrino signal 
 can be observed with a large statistical significance 
 via different flavor charged lepton final states. 

\thispagestyle{empty}

\newpage

\addtocounter{page}{-1}
\setcounter{footnote}{0}

\baselineskip 18pt

%%%%%%%%%%%%%%%%%%%%%%%%%%%%%%%%%
\section{Introduction}
%%%%%%%%%%%%%%%%%%%%%%%%%%%%%%%%%
The current experimental results on the neutrino oscillation 
 phenomena~\cite{PDG}, including the recent measurements of  
 the so-called reactor angle~\cite{T2K, MINOS, DCHOOZ, DayaBay, RENO}, 
 have established the existence of neutrino masses 
 and flavor mixings, 
 which require us to extend the Standard Model (SM). 
The seesaw extension of the SM~\cite{Seesaw} 
 is probably the simplest idea for explaining 
 the very small neutrino masses naturally, 
 where the SM-singlet heavy right-handed Majorana neutrinos 
 induce the dimension five operators leading to 
 very small light Majorana neutrino masses 
 (the seesaw mechanism~\cite{Seesaw}). 
The seesaw scale varies from the intermediate scale 
 to the electroweak scale 
 as we change the neutrino Dirac Yukawa coupling ($Y_D$)
 from the scale of top quark Yukawa coupling ($Y_D \sim 1$) 
 to the scale of electron Yukawa coupling ($Y_D \sim 10^{-6}$).

In high energy collider experimental point of view, 
 it is interesting if the heavy neutrino mass lies 
 at the TeV scale or smaller, because such heavy neutrinos 
 could be produced at high energy colliders, 
 such as the Large Hadron Collider (LHC) and 
 the International Linear Collider (ILC) 
 being projected as energy frontier physics in the future. 
However, since the heavy neutrinos are singlet 
 under the SM gauge group, they obtain the couplings 
 with the weak gauge bosons only through the mixing 
 via the Dirac Yukawa coupling. 
For the seesaw mechanism at the TeV scale or smaller, 
 the Dirac Yukawa coupling is too small 
 ($Y_D \sim 10^{-6}-10^{-5}$) to produce the observable 
 amount of the heavy neutrinos at the colliders.

There is another type of seesaw mechanism 
 so-called the inverse seesaw~\cite{InvSeesaw}, 
 where the small neutrino mass is obtained 
 by tiny lepton-number-violating parameters, 
 rather than the suppression by the heavy neutrino mass scale 
 in the ordinary seesaw mechanism. 
In the inverse seesaw scenario, the heavy neutrinos are 
 pseudo-Dirac particles and their Dirac Yukawa couplings 
 with the SM lepton doublets and the Higgs doublet 
 can be even order one, while reproducing the small light 
 neutrino masses. 
Thus, the heavy neutrinos in the inverse seesaw scenario 
 can be produced at the high energy colliders 
 through the sizable mixing with the SM neutrinos.

In this paper, we study the inverse seesaw scenario 
 and the heavy neutrino signatures at the LHC and ILC. 
For the concreteness of our inverse seesaw scenario 
 as well as the stability of the electroweak scale, 
 we consider a simple realization of the inverse seesaw mechanism 
 in the context of the next-to-minimal supersymmetric SM (NMSSM) 
 proposed in \cite{GOS}. 
The model parameters are fixed so as to reproduce 
 the neutrino oscillation data as well as other experimental 
 constraints such as precision measurements 
 of the weak gauge boson decays and 
 lepton-flavor-violating decays of charged leptons. 
We consider two typical cases in fitting neutrino oscillation data. 
One is that the flavor structure among light neutrinos 
 originates from the flavor structure of the neutrino 
 Dirac Yukawa couplings. 
In the other case, the Dirac Yukawa couplings are flavor-blind,
 and the flavor structure among light neutrinos is provided 
 by the lepton-number-violating parameters. 
Assuming the TeV scale mass for sparticles, 
 we concentrate on the production of the heavy neutrinos 
 with mass of ${\cal O}(100)$ GeV at the LHC and ILC, 
 and calculate the number of signal events. 
The heavy neutrino signals depend on the origin of 
 the flavor structure in the model and the types 
 of the hierarchical light neutrino mass spectra. 
We find that in some cases, the signal of the heavy neutrino 
 productions can be observed in the future collider experiments 
 with a large statistical significance 
 for the final states with different charged lepton flavors.

This paper is organized as follows. 
In Sec.~2, we introduce a model for the inverse seesaw 
 in the context of NMSSM. 
In Sec.~3 we give the explicit formulas of the heavy neutrino 
 production cross sections at the LHC and ILC 
 and of the partial decay widths, which are used in our numerical analysis. 
In Sec.~4, we first fix the model parameters 
 in simple parameterizations 
 so as to satisfy the experimental constraints, 
 assuming two typical flavor structures of our model  
 and two types of hierarchical neutrino mass spectra. 
For completeness, we also consider 
 a general parameterization of the neutrino Dirac 
 mass matrix. 
In Sec~5, the signal of the heavy neutrinos 
 at the LHC and ILC are investigated. 
For the general parameterization, we perform 
 parameter scan to identify the parameter region 
 to satisfy all experimental constraints, 
 for which we examine how much the heavy neutrino signal 
 is enhanced. 
Sec.~6 is devoted for conclusions.

%%%%%%%%%%%%%%%%%%%%%%%%%%%%%%%%%%%%%%%%%%%%%%%%%%%%%%
\section{Inverse Seesaw in NMSSM}
%%%%%%%%%%%%%%%%%%%%%%%%%%%%%%%%%%%%%%%%%%%%%%%%%%%%%%
As a simple realization of the inverse seesaw mechanism, 
 we consider an extension of the NMSSM \cite{NMSSM}. 
In the NMSSM, we introduce one gauge singlet chiral superfield $S$ 
 with even $Z_{2}$ matter parity 
 through the following superpotential terms: 
\bea
W\supset \lambda S H_{u} H_{d}+ \frac{\kappa}{3} S^{3}
\eea
 where $\lambda$ and $\kappa$ are the dimensionless constants, 
 $H_{u}$, $H_{d}$ are the MSSM Higgs doublets. 
We assume that the hidden sector breaks supersymmetry (SUSY) 
 and induces soft SUSY breaking terms for the MSSM scalers 
 and gauginos at the TeV scale, 
 by which the scalar $S$ and the MSSM Higgs doublets 
 develop the vacuum expectation values (VEVs). 
The VEV of $S$ generates the MSSM $\mu$ term: 
 $\mu = \lambda \langle S \rangle$.

%%%%%%%%%%%%%%%%%%%%%%%%%%%%%%%%%%%%%%%%%%%%%%%%%%%%%%
\begin{table}[ht]
\begin{center}
\begin{tabular}{c| c c|c c}
      &  SU(2)  & U(1)$_Y$ & $Z_3$ & $Z_2$ \\ 
\hline
$L$   & {\bf 2} & $-1/2$   & $ 1 $ & $-$   \\
$E^c$ & {\bf 1} & $ +1$    & $ \omega^2 $ & $-$   \\
$H_u$ & {\bf 2} & $+1/2$   & $ \omega $ & $+$   \\
$H_d$ & {\bf 2} & $-1/2$   & $ \omega $ & $+$   \\
$S$   & {\bf 1} & $0$      & $ \omega $ & $+$   \\ 
\hline
$N_j^c$ & {\bf 1} & $0$   & $ \omega^2 $ & $-$  \\
$N_j$   & {\bf 1} & $0$   & $ \omega $   & $-$  \\
\hline
\end{tabular}
\end{center}
\caption{ 
 The charge assignments of the NMSSM superfields. 
} \label{tab:11}
\end{table}
%%%%%%%%%%%%%%%%%%%%%%%%%%%%%%%%%%%%%%%%%%%%%%%%%%%%%%

We extend the NMSSM by introducing new particles 
 and a discrete $Z_3$ symmetry \cite{GOS}. 
The charge assignments for particles relevant to 
 the inverse seesaw mechanism are shown in Table~1. 
Here, $N_j^c$ and $N_j$ are the MSSM singlet particles, 
 heavy neutrinos of $j$-th generation, 
 and $\omega = e^{i 2 \pi/3}$. 
There are several possibilities for the $Z_3$ charge assignments, 
 and see \cite{GOS} for complete lists.

The renormalizable superpotential involving the new particles 
 and being consistent with all the symmetries is given by 
\bea
 W \supset  Y_{ij} L_i H_u N^c_j 
    + \frac{(\lambda_N)_{ij}}{2} S N_i N_j
    + m_{ii} N_i N^c_i.  
\eea
Without loss of generality, we have worked out in the basis  
 where the charged lepton Yukawa matrix and $m$ 
 are diagonalized. 
When the VEVs of $S$ and $H_u$ are developed,  
 we rewrite the superpotential as 
\bea
 W \supset  \nu^T m_D N^c + \frac{1}{2} N^T \mu N + N^T m  N^c,   
\eea 
 where we have used the matrix notation for generation indeces, 
 $\nu$ is the MSSM neutrino chiral superfield, 
 $ m_D = Y v \sin \beta/\sqrt{2}$ with $v=246$ GeV 
 is the neutrino Dirac mass matrix, 
 and $ \mu = \lambda_N \langle S \rangle$. 
For $m$  larger than the electroweak scale, 
 we integrate out the heavy fields  $N^c$ and $N$ 
 under the SUSY vacuum conditions,
\bea
&&  \frac{\partial W}{\partial N}=0   \; \to \;
  N^c = - m^{-1} \mu  N,   \nonumber \\
&&  \frac{\partial W}{\partial N^c}=0 \; \to \;
  N = - (m_D m^{-1})^T \nu ,
\label{SUSYvac} \eea
 and we arrive at the effective superpotential at low energies, 
\bea
  W_{\rm eff}= \frac{1}{2}
     \nu^T 
  \left[ 
     (m_D m^{-1}) \mu (m_D m^{-1})^T 
  \right] \nu . 
\eea
Note that the light Majorana neutrino mass matrix, 
 $m_\nu = (m_D m^{-1}) \mu (m_D m^{-1})^T $, 
 is proportional to $\mu$, so that tiny neutrino masses 
 can be realized by a small $\mu$ 
 even for both $m$ and $m_D$ being the electroweak scale. 
This is the inverse seesaw mechanism, 
 where the tiny neutrino mass corresponds to the breaking 
 of the lepton number by the tiny $\mu$ values.

Note that the heavy fields being integrated out 
 also have an impact on the  K\"ahler potential. 
Substituting the SUSY vacuum conditions into 
 the canonical K\"ahler potential for the heavy fields, 
 $\int d^4 \theta (N^\dagger N + N^{c \dagger} N^c)$,
 we obtain 
\bea
 {\cal K}_{\rm eff} = 
 \nu^\dagger \left[
  (m_D m^{-1})^* (m_D m^{-1})^T 
  \right] \nu + \cdots,     
\label{Eq-fig2} \eea
where the ellipsis  denote  higher order terms. 
Following the electroweak symmetry breaking, 
 this dimension six operator induces flavor-dependent corrections 
 to the kinetic terms of the left-handed neutrinos~\cite{dim6I}.

Assuming $m_D m^{-1} \ll 1$, 
 we can express the flavor eigenstates ($\nu$) of 
 the light Majorana neutrinos in terms of 
 the mass eigenstates of the light ($\nu_m$) 
 and heavy ($N_m$) Majorana neutrinos such as 
\bea 
  \nu \simeq {\cal N} \nu_m  + {\cal R} N_m,  
\eea 
where 
\bea
 {\cal R} = m_D m^{-1}, \; 
 {\cal N} =  \left(1 - \frac{1}{2} \epsilon \right) U_{\rm MNS} 
\eea
 with $\epsilon = {\cal R}^* {\cal R}^T$, 
 and $U_{MNS}$ is the usual neutrino mixing matrix 
 by which the mass matrix $m_\nu$ is diagonalized as  
\bea
   U_{MNS}^T m_\nu U_{MNS} = {\rm diag}(m_1, m_2, m_3). 
\eea
In the presence of $\epsilon$, the mixing matrix ${\cal N}$ 
 is not unitary. 
Using the mass eigenstates, the charged current interaction 
 in the Standard Model is given by 
\bea 
\mathcal{L}_{CC}= 
 -\frac{g}{\sqrt{2}} W_{\mu}
  \bar{e} \gamma^{\mu} P_L 
   \left( {\cal N} \nu_m+ {\cal R} N_m \right) + h.c., 
\label{CC}
\eea
where $e$ denotes the three generations of the charged 
 leptons in the vector form, and 
$P_L =\frac{1}{2} (1- \gamma_5)$ is the projection operator. 
Similarly, the neutral current interaction is given by 
\bea 
\mathcal{L}_{NC}= 
 -\frac{g}{2 c_w}  Z_{\mu} 
\left[ 
  \overline{\nu_m} \gamma^{\mu} P_L ({\cal N}^\dagger {\cal N}) \nu_m 
 +  \overline{N_m} \gamma^{\mu} P_L ({\cal R}^\dagger {\cal R}) N_m 
+ \left\{ 
  \overline{\nu_m} \gamma^{\mu} P_L ({\cal N}^\dagger  {\cal R}) N_m 
  + h.c. \right\} 
\right] , 
\label{NC}
\eea
 where $c_w=\cos \theta_w$ is the weak mixing angle. 
Because of non-unitarity of the matrix ${\cal N}$, 
 ${\cal N}^\dagger {\cal N} \neq 1$ and  
 the flavor-changing neutral current occurs.

%%%%%%%%%%%%%%%%%%%%%%%%%%%%%%%%%%%%%%%%%%%%%%%
\section{Productions and decays of heavy neutrinos at colliders}
%%%%%%%%%%%%%%%%%%%%%%%%%%%%%%%%%%%%%%%%%%%%%%%
In the previous section, we have found the charged and 
 neutral current interactions involving the heavy neutrinos. 
For detailed analysis, we need the information of 
 the mixing matrices, ${\cal N}$ and ${\cal R}$. 
In the next section, we will fix all the elements 
 of the matrices by considering the current experimental results. 
Before the analysis for fixing the parameters, 
 in this section we give the formulas 
 for the production cross sections and the partial decay widths 
 of the heavy neutrinos in the limit of 
 one generation and ${\cal N}={\cal R}=1$.

%%%%%%%%%%%%%%%%%%%%%%%%%%%%%%%%%%%%%%%%%%%%%%%
\subsection{Production cross section at LHC}
%%%%%%%%%%%%%%%%%%%%%%%%%%%%%%%%%%%%%%%%%%%%%%%
At the LHC, the heavy neutrinos can be produced 
 through the charged current interactions 
 by the $s$-channel exchange of the W boson. 
The main production process at the parton level is 
  $u \bar{d}\rightarrow e^{+} N_1$ 
 (and $\bar{u} d \rightarrow e^{-} \overline{N_1}$) 
 and the differential cross section is found to be 
\bea
\frac{d \hat{\sigma}_{LHC}}{d \cos\theta}
=
(3.89 \times 10^8 \;{\rm pb}) \times 
 \frac{\beta}{32 \pi \hat{s}} \frac{\hat{s} +M^2}{\hat s}
 \left( \frac{1}{2} \right)^2 3 \left( \frac{1}{3} \right)^2 
\frac{g^4}{4} 
\frac{({\hat s}^2-M^4)(2 + \beta \cos^2 \theta)}
{({\hat s} - m_W^2)^2+ m_W^2 \Gamma_W^2}, 
\eea
 where $\sqrt{\hat s}$ is the center-of-mass energy 
 of the colliding partons, $M$ is the mass of $N_1$, 
 and $\beta =({\hat s}-M^2)/({\hat s}+M^2)$.

The total production cross section at the LHC is given by 
\bea 
\sigma_{LHC} &=&
\int d \sqrt{\hat s} \int d \cos \theta 
\int^1_{{\hat s}/E_{CMS}^2} dx 
\frac{4 {\hat s}}{x E_{CMS}^2} 
f_u(x,Q) f_{\bar d}\left( \frac{\hat s}{x E_{CMS}},Q \right)  
\frac{d \hat{\sigma}_{LHC}}{d \cos\theta} \nonumber \\
&+& 
(u \to {\bar u}, {\bar d} \to d), 
\label{XLHC}
\eea 
where we have taken $E_{CMS} = 14$ TeV 
 for the center-of-mass energy of the LHC. 
In our numerical analysis, we employ CTEQ5M \cite{CTEQ} 
 for the parton distribution functions 
 for $u$-quark ($f_u$) and ${\bar d}$-quark ($f_{\bar d}$) 
 with the factorization scale $Q = \sqrt{\hat s}$. 
The total cross section as a function of $M$ 
 is depicted in Fig.~1.  
Since we have fixed ${\cal N}={\cal R}= 1$ in this analysis, 
 the resultant cross section shown in Fig.~1 
 corresponds to the maximum values for a fixed $M$.

%%%%%%%%%%%%%%%%%%%%%%%%%%%%%%%%%%%%%%%%%%%%%%%%%%
\begin{figure}[ht]
\begin{center}
\scalebox{1.5}{\includegraphics{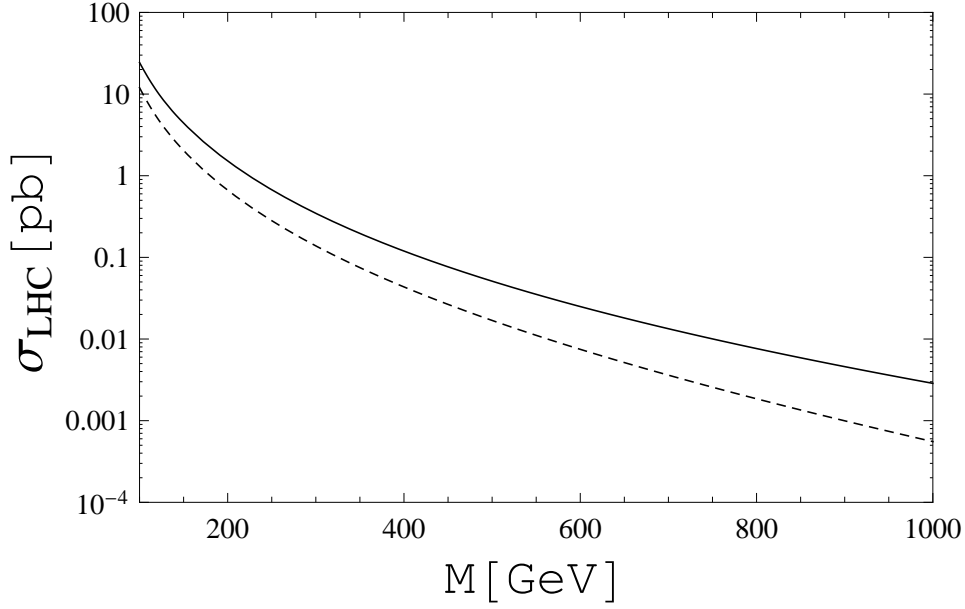}}
\end{center}
\caption{
The total production cross section of the heavy neutrino 
 as a function of its mass at the LHC with $\sqrt{s}=14$ TeV 
 (solid line). 
As a reference, the production cross section 
 at the LHC with $\sqrt{s}=8$ TeV is also plotted (dashed line).
}
\end{figure}
%%%%%%%%%%%%%%%%%%%%%%%%%%%%%%%%%%%%%%%%%%%%%%%%%%

There are three main modes for the heavy neutrino decays: 
 $N_1 \to e^- W^+$, $\nu_1 Z$, $\nu_1 h$. 
The corresponding partial decay widths are respectively given by
\bea
\Gamma(N_1 \rightarrow e^- W^+) 
 &=& \frac{g^2}{64 \pi} 
 \frac{ (M^2 - m_W^2)^2 (M^2+2 m_W^2)}{M^3 m_W^2} ,
\nonumber \\
\Gamma(N_1 \rightarrow \nu_1 Z) 
 &=& \frac{g^2}{128 \pi c_w^2} 
 \frac{ (M^2 - m_Z^2)^2 (M^2+2 m_Z^2)}{M^3 m_Z^2} ,
\nonumber \\
\Gamma(N_1 \rightarrow \nu_1 h) 
 &=& \frac{(M^2-m_h^2)^2}{32 \pi M} 
 \left( \frac{1}{v \sin \beta}\right)^2  .
\label{widths}
\eea 
The long-sought Higgs boson is finally discovered by 
 the ATLAS~\cite{ATLAS} and the CMS~\cite{CMS} collaborations 
 at the LHC. 
According to the discovery, we use 
 $m_h=125.3$ GeV~\cite{CMS} in the following analysis. 
Our results are almost independent of the choice of 
 the Higgs boson mass in the range of $125-126$ GeV.

%%%%%%%%%%%%%%%%%%%%%%%%%%%%%%%%%%%%%%%%%%%%%%%
\subsection{Production cross section at ILC}
%%%%%%%%%%%%%%%%%%%%%%%%%%%%%%%%%%%%%%%%%%%%%%%
The ILC can produce the heavy neutrino 
 in the process $e^+ e^- \to \overline{\nu_1} N_1$ 
 through $t$ and $s$-channels exchanging the W and Z bosons, respectively. 
The total differential production cross section for this process 
 is calculated as 
\bea
%&& 
\frac{d \sigma_{ILC}}{d \cos\theta}  
&=& (3.89\times 10^{8} \; {\rm pb}) \times 
\frac{\beta}{32 \pi s} 
\frac{s + M^2}{s} 
\left( \frac{1}{2} \right)^2 
\nonumber \\
&\times& 
\left[   
\frac{16  C_1^2 C_2^2 \left( s^2 - M^4 \right) 
(1+\cos\theta) (1+ \beta \cos\theta)}
{(M^2 -\frac{s -M^2}{2} (1-\beta \cos\theta)- m_W^2)^{2}+ m_W^2 \Gamma_W^2} 
\right.
\nonumber \\
&+&
\frac{ \left( 
4 (C^{2}_{A_{e}}+C^{2}_{V_{e}})
(C^{2}_{A_{\nu}}+C^{2}_{V_{\nu}}) (1+\beta \cos^2 \theta )
+ 16 C_{A_{e}}C_{V_{e}}C_{A_{\nu}}C_{V_{\nu}} (1+\beta) \cos\theta 
\right)(s^2 - M^4)}{(s -m_Z^2)^2 + m_Z^2 \Gamma_Z^2} 
\nonumber \\
&-&
32 C^{2}_{1} C^{2}_{A_{e}} (s^2 - M^4 ) 
(1+ \cos\theta)(1+\beta \cos\theta) 
\nonumber \\
&&
\left.
 \times 
\frac{
\left(M^2 - \frac{s-M^2}{2}(1-\beta \cos\theta) - m_W^2 \right)(s - m_Z^2) 
+ m_W m_Z \Gamma_W \Gamma_Z  
}
{
((M^2 - \frac{s-M^2}{2} (1-\beta \cos\theta) -m_W^2)^{2} + m_W^2 \Gamma_W^2 )
 ( (s - m_Z^2)^2 + m_Z^2 \Gamma_Z^2 ) 
}
\right], 
\label{XILC}
\eea
where $\beta=(s-M^2)/(s+M^2)$,  
\bea
&&
C_{1}= -C_2 = \frac{g}{2 \sqrt{2}}, \; 
C_{A_{\nu}}= C_{V_{\nu}}= \frac{g}{4 \cos \theta_W}, 
\nonumber \\ 
&&
C_{A_{e}}= 
\frac{g}{2 \cos \theta_w} 
\left( 
-\frac{1}{2} + 2 \sin^2 \theta_w 
\right), \; 
C_{V_{e}}= - \frac{g}{4 \cos\theta_w}.
\eea
The total production cross section for the process 
 $e^+ e^- \to \overline{\nu_1} N_1$ 
 for the ILC with $\sqrt{s}=500$ GeV and 1 TeV, 
 respectively, are shown in Fig.~2. 
Since we have fixed ${\cal N}={\cal R}=1$ 
 in this analysis, the resultant cross section shown in Fig.~2 
 corresponds to the maximum values for a fixed $M$.

%%%%%%%%%%%%%%%%%%%%%%%%%%%%%%%%%%%%%%%%%%%%%%%%%%
\begin{figure}[ht]
\begin{center}
\scalebox{1.5}{\includegraphics{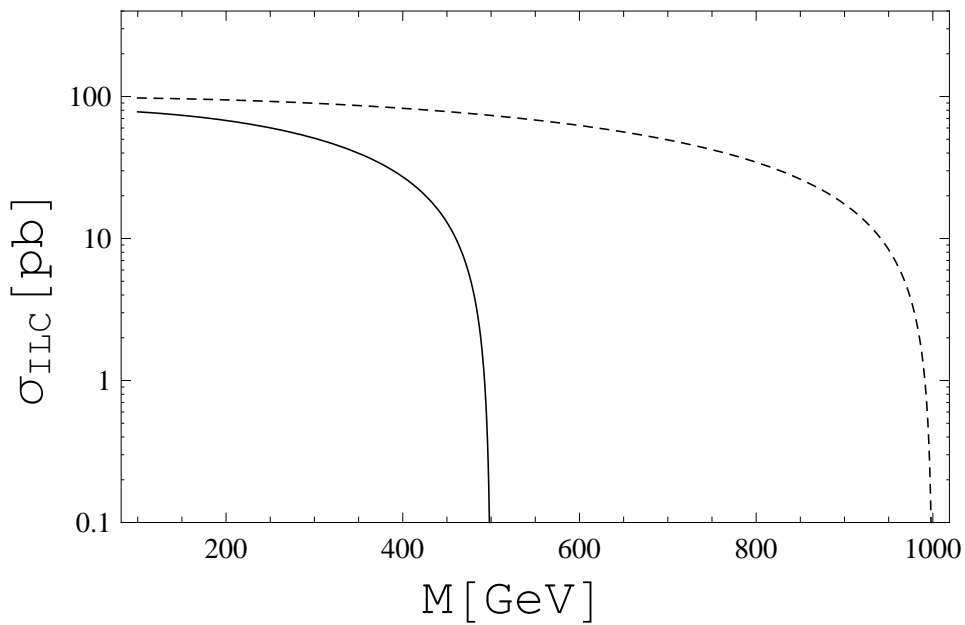}}
\end{center}
\caption{
The total production cross section of the process
 $e^+ e^- \rightarrow \overline{\nu}_1 N_1$ 
 at the ILC with $\sqrt{s}=500$ GeV (solid line) 
 and $\sqrt{s}=1$ TeV (dashed line).
}
\end{figure}
%%%%%%%%%%%%%%%%%%%%%%%%%%%%%%%%%%%%%%%%%%%%%%%%%%

%%%%%%%%%%%%%%%%%%%%%%%%%%%%%%%%%%%%%%%%%%%%%%%%%%%%%%%%%%
\section{Fixing the matrices ${\cal N}$ and ${\cal R}$} 
%%%%%%%%%%%%%%%%%%%%%%%%%%%%%%%%%%%%%%%%%%%%%%%%%%%%%%%%%%
\subsection{Simple parameterizations} 
%%%%%%%%%%%%%%%%%%%%%%%%%%%%%%%%%%%%%%%%%%%
The elements of the matrices ${\cal N}$ and ${\cal R}$ 
 are constrained by the experimental 
 results~\cite{Constraints1, Constraints2, Constraints3}. 
We begin with the current neutrino oscillation data. 
Recently non-zero reactor neutrino angle $\theta_{13}$ 
 has been observed in several experiments, 
 such as T2K~\cite{T2K}, MINOS~\cite{MINOS}, 
 Double CHOOZ~\cite{DCHOOZ}, Daya Bay~\cite{DayaBay} 
 and RENO~\cite{RENO}, 
 and their results are consistent with each other. 
Together with other oscillation data, 
 all neutrino oscillation parameters except the Dirac CP-phase, 
 two mass squared differences and three mixing angles 
 have been measured in some precision. 
By using the data, we fix the neutrino mixing matrix elements. 
In the following analysis, 
 we adopt $\sin^{2}2{\theta_{13}}=0.092$~\cite{DayaBay} 
 along with the other oscillation data: 
 $\sin^2 2\theta_{12}=0.87$, $\sin^2 2\theta_{23}=1.0$, 
 $\Delta m_{12}^2 = m_2^2-m_1^2 = 7.6 \times 10^{-5}$ eV$^2$, 
 and $\Delta m_{23}^2= |m_3^2-m_2^2|=2.4 \times 10^{-3}$ eV$^2$. 
Then, the numerical values of neutrino mixing matrix elements 
 are explicitly given by 
\bea
U_{MNS}=\begin{pmatrix}
 0.815 &  0.559 & 0.153 \\
-0.489 &  0.522 & 0.699 \\ 
 0.310 & -0.645 & 0.699 
\end{pmatrix}, 
\label{Umns}
\eea
where we have fixed all the CP-phases to be zero, for simplicity. 
We will discuss a general parameterization 
 including all CP-phases as well as an arbitrary orthogonal 
 matrix in the next subsection.

For the neutrino mass spectrum, we consider both 
 the normal hierarchy (NH) and the inverted hierarchy (IH).  
The lightest mass eigenstate is assumed to be very light 
 and its mass is approximated as 0. 
Thus, in the NH case, the diagonal mass matrix is given by 
\bea 
  D_{NH} ={\rm diag}
  \left(0, \sqrt{\Delta m_{12}^2},
           \sqrt{\Delta m_{12}^2 + \Delta m_{23}^2} \right),  
\label{DNH}
\eea 
while in the IH case 
\bea 
  D_{IH} ={\rm diag}
\left( \sqrt{\Delta m_{23}^2 - \Delta m_{12}^2}, 
 \sqrt{\Delta m_{23}^2}, 0 \right).  
\label{DIH}
\eea

In order to make our discussion simple, 
 we assume the degeneracy of the heavy neutrinos in mass 
 such as $ m = M {\bf 1}$ with the unit matrix ${\bf 1}$, 
 so that the neutrino mass matrix is simplified as 
\bea 
 m_\nu = {\cal R} \mu {\cal R}^T 
       = \frac{1}{M^2} m_D \mu m_D^T. 
\eea 
Moreover we consider two typical cases for the flavor 
 structure of the model: 
(i) $\mu$ is also proportional to the unit matrix, 
    $\mu \to \mu {\bf 1}$. 
In this case, the flavor structure of $m_\nu$ 
 is provided by a non-diagonal $m_D$. 
We call this case Flavor Non-Diagonal (FND) case. 
(ii) The other case is what we call Flavor Diagonal (FD) case,
 where $m_D$ is proportional  to the unit matrix, $m_D \to m_D {\bf 1}$ 
 and thus the flavor structure is encoded 
 in the $3 \times 3$ matrix $\mu$.

In the FND case, we consider two generations of $N_j$ and $N^c_j$ 
 with $j=1,2$, so that 
\bea 
   m_\nu = \frac{\mu}{M^2} m_D m_D^T 
 = U_{MNS}^* D_{NH/IH} U_{MNS}^\dagger.  
\eea  
 From this formula, we parameterize the neutrino Dirac mass matrix as 
\bea 
  m_D = \frac{M}{\sqrt{\mu}} U_{MNS}^* \sqrt{D_{NH/IH}},  
\label{mD}
\eea
where the matrices denoted as $\sqrt{D_{NH/IH}}$ are defined as 
\bea
\sqrt{D_{NH}} =
\begin{pmatrix}
    0 & 0 \\
 (\Delta m_{12}^2)^{\frac{1}{4}} & 0 \\
             0   & (\Delta m_{23}^2+\Delta m_{12}^2)^{\frac{1}{4}} \\
\end{pmatrix},  \; \; 
\sqrt{D_{IH}} =
\begin{pmatrix}

    (\Delta m_{23}^2 - \Delta m_{12}^2)^{\frac{1}{4}} & 0 \\
    0 & (\Delta m_{23}^2)^{\frac{1}{4}} \\
    0 & 0 \\
\end{pmatrix}. 
\eea 
Note that in the case with two generations of $N_j$ and $N^c_j$, 
 the lightest mass eigenvalue is exactly 0. 
On the other hand, in the FD case, we have 
\bea 
 m_\nu =\left(\frac{m_D}{M} \right)^2 \mu 
 = U_{MNS}^* D_{NH/IH} U_{MNS}^\dagger.  
\eea

Due to its non-unitarity, 
 the elements of the mixing matrix ${\cal N}$ 
 are severely constrained by the combined data 
 from neutrino oscillation experiments, 
 the precision measurement of weak boson decays, 
 and the lepton-flavor-violating decays 
 of charged leptons~\cite{Constraints1, Constraints2, Constraints3}. 
We update the results by using more recent data 
 on the lepton-favor-violating decays~\cite{Adam, Aubert, OLeary}: 
\bea
|{\cal N}{\cal N}^\dagger| =
\begin{pmatrix} 
 0.994\pm0.00625& 1.499 \times 10^{-5} &  8.764\times 10^{-3}\\
 1.499 \times 10^{-5} & 0.995\pm 0.00625 & 1.046\times 10^{-2}\\
 8.764 \times 10^{-3}& 1.046 \times 10^{-2} & 0.995\pm 0.00625
\end{pmatrix} .
\eea
Since ${\cal N}{\cal N}^\dagger \simeq {\bf 1} - \epsilon$, 
 we have the constraints on $\epsilon$ such that 
\bea
|\epsilon| =
\begin{pmatrix} 
 0.006\pm0.00625& < 1.499 \times 10^{-5} & < 8.764 \times 10^{-3}\\
 < 1.5\times 10^{-5} & 0.005\pm 0.00625 & < 1.046 \times 10^{-2}\\
 < 8.76356\times 10^{-3}& < 1.046 \times 10^{-2} & 0.005\pm 0.00625
\end{pmatrix} .
\eea
The most stringent bound is given by the $(1,2)$ element 
 which is from the constraint on the lepton-flavor-violating 
 muon decay $\mu \to e \gamma$\footnote{
It has been pointed out~\cite{mue1, mue2} that 
 in the SUSY inverse seesaw model, 
 sparticle Z-penguin contributions can dominate 
 the lepton-flavor-violating processes,  
 independently of sparticle mass spectrum. 
According to the analysis in Ref.~\cite{mue2}, 
 we have found that the constraint from $\mu-e$ conversion process 
 is more severe than the one from the $\mu \to e \gamma$ process 
 for $M > 335$ GeV. 
Since we will focus on $M=100-150$ GeV in the following analysis, 
 we use the value of $\mu_{min}$ determined 
 from the muon decay constraints. 
}. 
For the FND case, we describe $\epsilon$ as 
\bea 
 \epsilon = \frac{1}{M^2} m_D m_D^T 
          = \frac{1}{\mu} U_{MNS} D_{NH/IH} U_{MNS}^T,  
\eea 
and determine the minimum $\mu$ value ($\mu_{min}$) 
 so as to give $\epsilon_{1 2}=1.5 \times 10^{-5}$ 
 using the oscillation data 
 in Eqs.~(\ref{Umns}), (\ref{DNH}) and (\ref{DIH}). 
We have found $\mu_{min}=525$ eV and $329$ eV 
 for the NH and IH cases, respectively. 
Here we have used the fact that all parameters are real 
 according to our assumption. 
In this way, we can completely determine 
 the mixing matrix ${\cal R}$ and ${\cal N}$ from Eq.~(\ref{mD}) 
 by taking $\mu = \mu_{\min}$, which optimizes 
 the production cross sections of the heavy neutrinos 
 at the LHC and ILC. 
For the FD case, we simply take  
 $\epsilon = (m_D/M)^2 {\bf 1}=0.01225 {\bf 1}$ ($95.5\%$ CL).

%%%%%%%%%%%%%%%%%%%%%%%%%%%%%%%%%%%%%%%%%%%%
\subsection{General parameterization}  
%%%%%%%%%%%%%%%%%%%%%%%%%%%%%%%%%%%%%%%%%%%%
For completeness, we also consider a general parameterization  
 for the neutrino Dirac mass matrix for the FND case. 
From the inverse seesaw formula, 
\bea
 m_\nu = \mu {\cal R} {\cal R}^T 
   =\frac{\mu}{M^2}m_Dm_D^T= U_{MNS}^*D_{NH/IH} U_{MNS}^\dagger, 
\eea
we can generally parameterize ${\cal R}$ as 
\bea
{\cal R}(\delta,\rho,x,y)= 
\frac{1}{\sqrt{\mu}}U_{MNS}^*\sqrt{D_{NH/IH}}O ,
\eea
where $O$ is a general orthogonal matrix expressed as 
\bea 
O =
\begin{pmatrix}
    \cos \alpha & \sin \alpha \\
    -\sin \alpha & \cos\alpha\\
\end{pmatrix}
=
\begin{pmatrix}
    \cosh y& i \sinh y \\
   -i \sinh y & \cosh y\\
\end{pmatrix}
\begin{pmatrix}
    \cos x& \sin x \\
   -\sin x & \cosh x\\
\end{pmatrix},
\eea
with a complex number $\alpha=x +i y$, and 
the general form of the neutrino mixing matrix, 
\bea
U_{MNS}=
\begin{pmatrix}
C_{12} C_{13}&S_{12} C_{13}&S_{13} e^{i\delta}\\
-S_{12} C_{23}-C_{12} S_{23} S_{13} e^{i\delta}&C_{12} C_{23} - S_{12} S_{23} S_{13} e^{i\delta}&S_{23} C_{13}\\
S_{12} S_{23}-C_{12} C_{23} S_{13} e^{i\delta}&-C_{12} S_{23}-S_{12} C_{23} S_{13} e^{i\delta}&C_{23} C_{13}
\end{pmatrix}
\begin{pmatrix}
1&0&0\\
0&e^{i\rho}&0\\
0&0&1
\end{pmatrix}.
\eea
Here, $C_{ij}= \cos \theta_{ij}$, $S_{ij}=sin\theta_{ij}$, 
 $\delta$ is the Dirac phase and $\rho$ is the Majorana phase. 
Thus, in this general parameterization, we have 
\bea
\epsilon(\delta, \rho, y )= {\cal R}^*{\cal R}^T
   =\frac{1}{\mu} U_{MNS}\sqrt{D_{NH/IH}}O^{*}O^{T}
\sqrt{D_{NH/IH}}^{T}U_{MNS}^{\dagger}.  
\label{eps}
\eea
Note that 
\bea
O^{*}O^{T}=
\begin{pmatrix}
   \cosh^2 y+ \sinh^2 y&-2i \cosh y \sinh y \\
   2i \cosh y \sinh y & \cosh^2 y + \sinh^2 y
\end{pmatrix}
\eea
 is independent of $x$, and hence the $\epsilon$-matrix is 
 a function of $\delta$, $\rho$ and $y$.

In the next section, we perform a parameter scan 
 under the experimental constraints 
 and identify an allowed region for 
 the parameter set $\{\delta, \rho, y \}$. 
Then, we calculate the heavy neutrino production cross section 
 for the parameter set and examine how much 
 the production cross section is enhanced, 
 satisfying the experimental constrains.

%%%%%%%%%%%%%%%%%%%%%%%%%%%%%%%%%%%%%%%%%%%%%%%%%%%%%%%%%%%%%%%%%%%
\section{Collider signatures of heavy neutrinos}
%%%%%%%%%%%%%%%%%%%%%%%%%%%%%%%%%%%%%%%%%%%%%%%%%%%%%%%%%%%%%%%%%%%
Let us now investigate the collider signatures of the heavy neutrinos 
 with the information of ${\cal R}$ and ${\cal N}$ 
 determined by the previous sections. 
In Sec.~3, we have already given the formulas used in our analysis 
 in the limit of ${\cal R}={\cal N}=1$. 
It is easy to generalize the formulas 
 with the concrete ${\cal R}$ and ${\cal N}$. 
The production cross section of the $i$-th generation  
 heavy neutrino at the LHC, through the process 
 $q \bar{q}^\prime \to \ell N_i$  
 ($u \bar{d} \to \ell_\alpha^{+} N_i$ 
  and $ {\bar u} d \to \ell_\alpha^{-} \overline{N_i}$)
  is given by 
\bea 
 \sigma(q \bar{q}^\prime \to \ell_\alpha N_i) 
 = \sigma_{LHC} |{\cal R}_{\alpha i}|^2,  
\eea
where $\sigma_{LHC}$ is the cross section given in Eq.~(\ref{XLHC}). 
Similarly, the production cross section at the ILC is 
\bea 
 \sigma(e^+ e^- \to \overline{\nu_\alpha} N_i) 
 = \sigma_{ILC} |{\cal R}_{\alpha i}|^2,  
\eea
where $\sigma_{ILC}$ is given in Eq.~(\ref{XILC}), 
 and we have used the approximation 
 ${\cal N}^\dagger {\cal R} \simeq U_{MNS}^\dagger {\cal R}$ 
 because $|\epsilon_{\alpha \beta}| \ll 1$ 
 as discussed in the previous section. 
The partial decay widths for the process 
 $N_i \to \ell_\alpha^{-} W^+/\nu_\alpha Z/ \nu_\alpha h$ 
 are obtained by multiplying Eq.~(\ref{widths}) 
 and the factor $|{\cal R}_{\alpha i}|^2$ together.

%%%%%%%%%%%%%%%%%%%%%%%%%%%%%%%%%%%%%%%%%%%%%%%%%
\begin{figure}[t]
\begin{tabular}{cc}
\begin{minipage}{0.5\hsize}
\begin{center}
\includegraphics[scale=0.8]{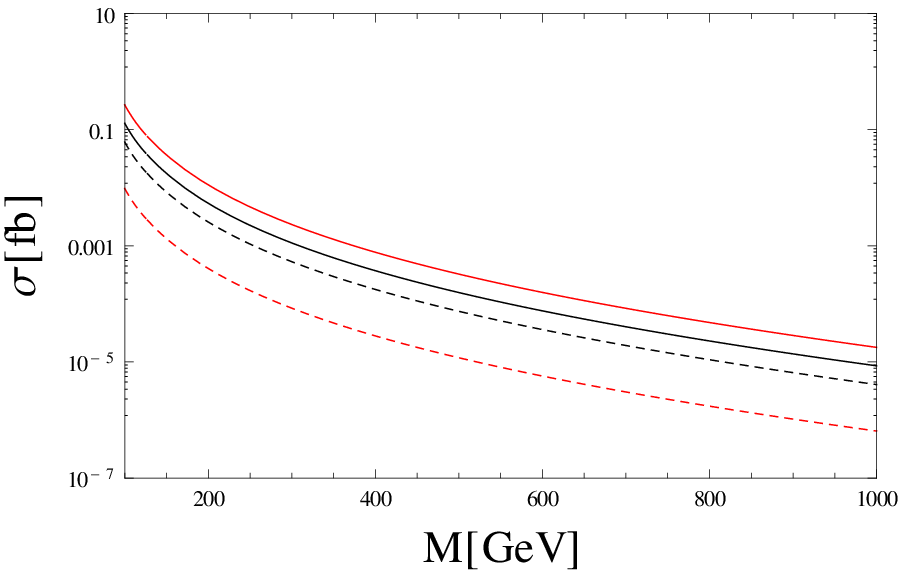}
\end{center}
\end{minipage}
\begin{minipage}{0.5\hsize}
\begin{center}
\includegraphics[scale=0.8]{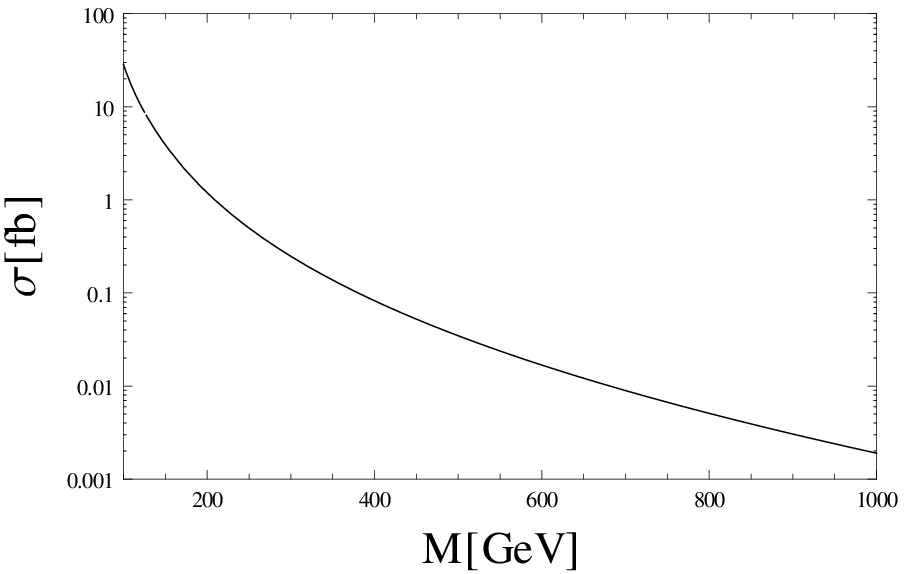}
\end{center}
\end{minipage}
\end{tabular}
\caption{
Signal cross sections providing the tri-lepton final states 
 for the FND (left panel) and FD (right panel) cases, 
 at the LHC with $\sqrt{s}=14$ TeV. 
}
%\label{DMLEP}
\end{figure}
%%%%%%%%%%%%%%%%%%%%%%%%%%%%%%%%%%%%%%%%%%%%%%%%%

%%%%%%%%%%%%%%%%%%%%%%%%%%%%%%%%%%%%%%%%%%%
\subsection{Heavy neutrino signal at LHC 
 with the simple parameterizations}
%%%%%%%%%%%%%%%%%%%%%%%%%%%%%%%%%%%%%%%%%%%
As has been studied in Ref.~\cite{tri-lepton} 
 (see also \cite{tri-lepton2} for the studies 
  on the left-right symmetric model), 
 the most promising signal of the heavy neutrino productions 
 at the LHC is obtained by the final state 
 with three charged leptons 
 ($ \ell^\pm \ell^\pm \ell^\mp$ with the total charge $\pm 1$) 
 through the process $q \bar{q}^\prime \to N \ell^{\pm}$ 
 followed by $ N \to \ell^{\pm} W^{\mp}$ 
 and $W^{\mp} \to \ell^{\mp} \nu$. 
In this work, detailed studies have been performed 
 for the signal of the heavy neutrino with a 100 GeV mass, 
 which couples with either the electron or the muon.  
The events were pre-selected for two like-sign charged leptons 
 ($e e$ or $\mu \mu$) to have transverse momentum $p_T > 30$ GeV. 
The decay mode, $ N \to \nu Z$, followed by 
 $Z \to \ell^+ \ell^-$ is rejected by a cut 
 for the invariant mass of the charge neutral di-lepton. 
After elaborate selections, it has been concluded \cite{tri-lepton} 
 that the heavy neutrino coupling to the muon could be observed 
 at the LHC through the tri-lepton final states.

In our analysis, we follow the procedure in \cite{tri-lepton}. 
Since we are considering the general case 
 with ${\cal R}$ and ${\cal N}$ consistent 
 with the updated experimental data, 
 the production cross sections of the heavy neutrinos 
 are different from the ones in \cite{tri-lepton}. 
Fig.~3 shows the signal cross section providing  
 tri-lepton final states with $e e$ or $\mu \mu$ 
 for the FND (left) and FD (right) cases, as a function of 
 the heavy neutrino mass. 
In the left panel, the dashed and solid lines correspond to 
 the NH and IH cases, respectively. 
The upper solid (dashed) line shows the cross sections 
 with $e e$ ($\mu \mu$).

We adopt the same efficiency for the signal events 
 and the SM background events which was found in \cite{tri-lepton}. 
The number of events for tri-lepton final states 
 with $ ee $ and $\mu \mu$, respectively, are listed on Table~2, 
 for the luminosity $30$ fb$^{-1}$. 
Unfortunately, the number of events for the FND case 
 are found to be too small. 
This is because the component of ${\cal R}$ is severely 
 constrained to be small by the current experiments. 
On the other hand, the FD case results the large number 
 of signal events specially in the $\mu\mu$ case with a significance 
 of more than 5-$\sigma$. 
If we naively estimate the significance by $S/\sqrt{B}$, 
 the luminosity of $25$ fb$^{-1}$ ($11$ fb$^{-1}$) 
 is required to achieve 5-$\sigma$ significance 
 for the ($\mu \mu$) final states.

%%%%%%%%%%%%%%%%%%%%%%%%%%%%%%%%%%%%%%%%%%%%%%%%%%%%%%%%%%%%%%%%%
\begin{table}[t]
\begin{center}
\begin{tabular}{c|cc}
\hline
         & $ ee $ & $\mu \mu$ \\ 
\hline 
FND (NH) & $ 0.254  $  & $ 1.61 $  \\
FND (IH) & $ 7.00   $  & $3.38$  \\
FD       & $58.7    $  & $56.2$  \\
\hline 
SM background    & $116.4$   & $45.6$   \\ 
\hline 
\end{tabular}
\end{center}
\caption{
Number of events at the LHC with $\sqrt{s}=14$ TeV 
 and $30$ fb$^{-1}$ luminosity, 
 for the heavy neutrino mass $M=100$ GeV.
}
\end{table}

%%%%%%%%%%%%%%%%%%%%%%%%%%%%%%%%%%%%%%%%%%%
\subsection{
Heavy neutrino signal at ILC 
 with the simple parameterizations
}
%%%%%%%%%%%%%%%%%%%%%%%%%%%%%%%%%%%%%%%%%%%

%%%%%%%%%%%%%%%%%%%%%%%%%%%%%%%%%%%%%%%%%%%%%%%%%
\begin{figure}[t]
\begin{center}
\includegraphics[scale=0.9]{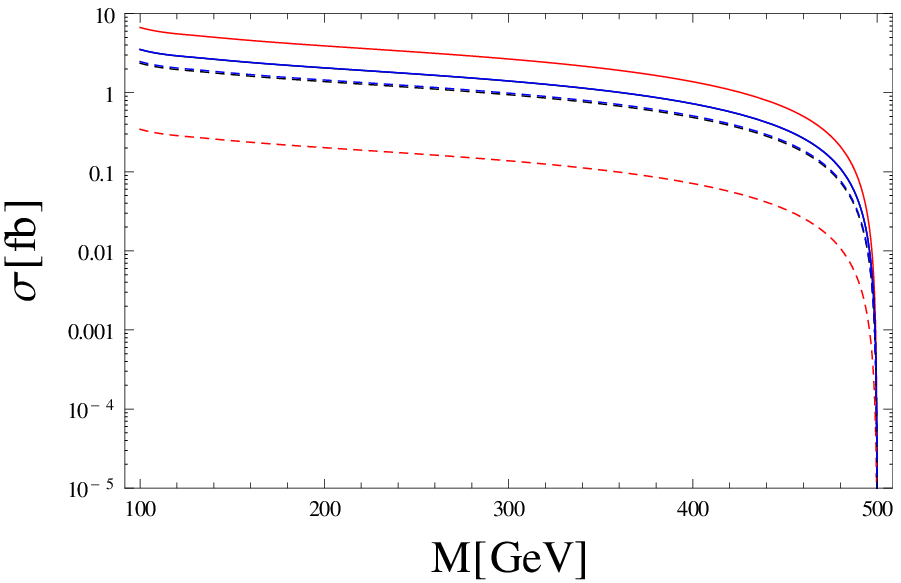}
\includegraphics[scale=0.9]{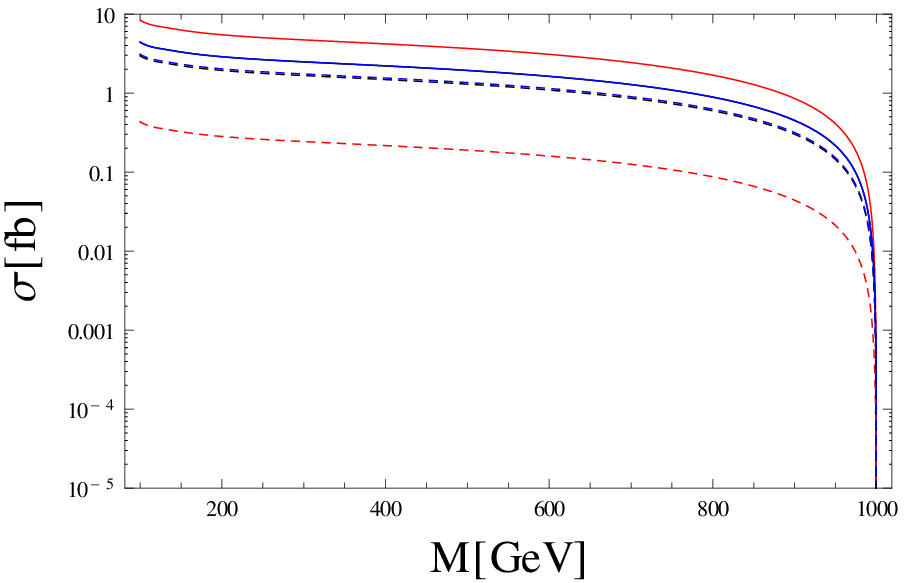}
\includegraphics[scale=0.9]{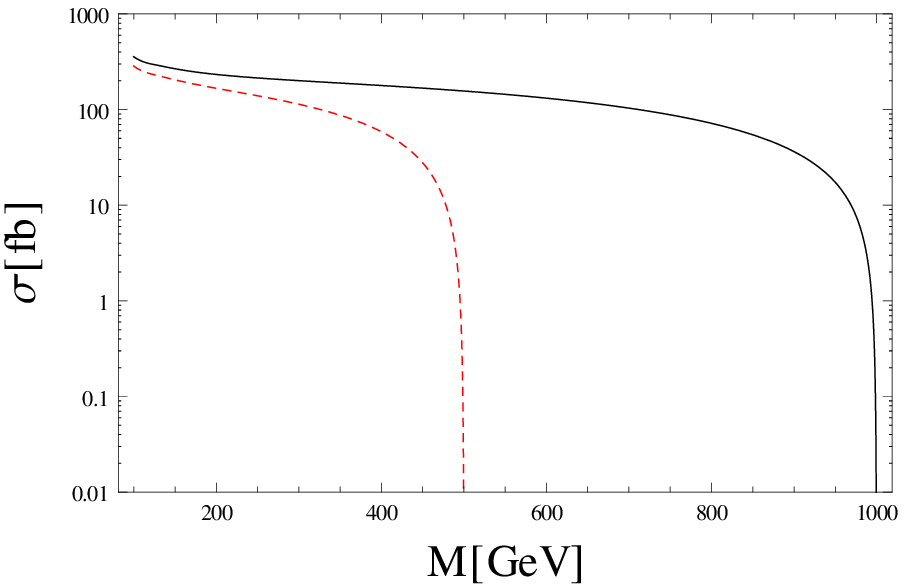}
\end{center}
\caption{
The production cross sections for the process 
 $e^+ e^- \to \nu N $, followed by the decays 
 $N \to \ell W$ ($\ell=e, \mu, \tau$) and $W \to q {\bar q}^\prime$, 
 as a function of the heavy neutrino mass. 
The upper-left panel shows the results for the FND case 
 with $\sqrt{s}=500$ GeV. 
The upper-right panel is the same as the upper-left panel 
 but for the case with $\sqrt{s}=1$ TeV. 
The results for the FD case are shown in the lower panel  
 for $\sqrt{s}=500$ GeV (solid) and $\sqrt{s}=1$ TeV (dashed),
 respectively. 
}
\label{DMDD}
\end{figure}
%%%%%%%%%%%%%%%%%%%%%%%%%%%%%%%%%%%%%%%%%%%%%%%%%

The signature of heavy neutrinos at the ILC 
 has been studied in detail based on the realistic Monte Carlo 
 simulations  in \cite{FujiiEtAl}. 
In the studies, a five-dimensional model with 
 bulk right-handed neutrinos \cite{XdimNu} is considered 
 and its 4-dimensional effective theory provides 
 the Kaluza-Klein tower of the heavy neutrinos 
 having sizable coupling to the weak gauge bosons 
 through mixings with the SM light neutrinos. 
This structure of the couplings between the heavy neutrinos 
 and the SM particles in the five-dimensional model 
 is similar to the one in our inverse seesaw model. 
Thus, we apply the results in \cite{FujiiEtAl}, 
 in particular, the signal and background selection procedure 
 to our model.

According to \cite{FujiiEtAl}, we focus on the two-jets and 
 one isolated lepton signal with large missing energy at the ILC: 
 $e^+e^- \to \nu N$, followed by the decays 
 $N \to \ell W$ and $W \to q {\bar q}^\prime$, 
 through which the heavy neutrino production cross sections 
 and the heavy neutrino mass can be reconstructed. 
The production cross sections for the process 
 $e^+ e^- \to \nu N $, followed by the decays 
 $N \to \ell W$ ($\ell=e, \mu, \tau$) and $W \to q \bar{q}^\prime$, 
 as a function of the heavy neutrino mass 
 are depicted in Fig.~4. 
The upper-left panel shows the results for the FND case 
 with $\sqrt{s}=500$ GeV. 
The dashed lines from top to bottom denote the signal cross sections 
 for $\ell=\tau$, $\mu$ and $e$, respectively, in the NH case. 
The two lines corresponding to $\ell=\tau$ and $\mu$ 
 are well-overlapping. 
The solid lines from top to bottom denote the signal cross sections 
 for $\ell=e$, $\tau$ and $\mu$, respectively, in the IH case. 
The two dashed lines corresponding to $\ell=\mu$ and $\tau$ 
 are well-overlapping. 
The upper-right panel is the same as the upper-left panel 
 but for the case with $\sqrt{s}=1$ TeV. 
The results for the FD case are shown in the lower panel  
 for $\sqrt{s}=500$ GeV (dashed) and $\sqrt{s}=1$ TeV (solid),
 respectively. 
Here $\ell$ is either $e$, $\mu$ or $\tau$.

For the ILC with $\sqrt{s}=500$ GeV and the luminosity 
 ${\cal L}=500$ fb$^{-1}$, 
 the signal and background events are listed on Table~2, 
Here, the final state of one electron and two jets 
 with missing energy from anti-neutrinos is considered, 
 and we have adopted the efficiencies found in \cite{FujiiEtAl}, 
 for $M=150$ GeV. 
The main backgrounds are 
 $e \nu W \to e \nu q \bar{q}$ and $WW \to \ell \nu q \bar{q}$, 
 which are dramatically reduced by the selection 
 using an isolated-electron track with 
 a requirement of its energy range, 10 GeV$\leq E_e \leq 200$ GeV, 
 the requirement of the reconstructed  di-jet mass to be 
 consistent with W hypothesis etc (see \cite{FujiiEtAl} for details).
The signal and background events 
 for $\sqrt{s}=1$ TeV and the same luminosity ${\cal L}=500$ fb$^{-1}$  
 are listed on Table~3. 
For completeness, we have also listed the signal events (without cuts) 
 for the case of $\ell=\mu$ and $\tau$. 

%%%%%%%%%%%%%%%%%%%%%%%%%%%%%%%%%%%%%%%%%%%%%%%%%%%%%%
\begin{table}[ht]
\begin{center}
\begin{tabular}{c|c|c}
           & Events before cuts & Events after cuts \\ 
\hline
FND (NH)   &  123.7    &  84.04   \\ 
FND (IH)   &  2397     &  1363   \\ 
FD         &  102210    &  69189.7   \\ 
\hline
SM background      & 3210500    &  23346  \\ 
\hline
\hline
FND (NH, $\ell=\mu $)  &  847.5     &    \\ 
FND (NH, $\ell=\tau$)  &  887.0    &    \\ 
\hline
FND (IH, $\ell=\mu $)  &  1261     &    \\ 
FND (IH, $\ell=\tau$)  &  1266     &    \\ 
\hline
\end{tabular}
\end{center}
\caption{ 
 The number of events at the ILC with $\sqrt{s}=500$ GeV 
 and the luminosity 500 fb$^{-1}$, 
 for the heavy neutrinos with mass 150 GeV. 
We have adopted the efficiencies for the signal and the SM background 
 found by the realistic Monte Carlo simulations in \cite{FujiiEtAl}. 
} 
%\label{tab:2}
\end{table}
%%%%%%%%%%%%%%%%%%%%%%%%%%%%%%%%%%%%%%%%%%%%%%%%%%%%%%

%%%%%%%%%%%%%%%%%%%%%%%%%%%%%%%%%%%%%%%%%%%%%%%%%%%%%%
\begin{table}[ht]
\begin{center}
\begin{tabular}{c|c|c}
           & Events before cuts & Events after cuts \\ 
\hline
FND (NH)   &  162    &    52.0   \\ 
FND (IH)   &  3133    &   776.1   \\ 
FD         &  133605    &  42671.3   \\ 
\hline
SM background     & 5476408  &  10500  \\ 
\hline
\hline
FND (NH, $\ell=\mu $)  &  1108    &    \\ 
FND (NH, $\ell=\tau$)  &  1160    &    \\ 
\hline
FND (IH, $\ell=\mu $)  &  1648    &    \\ 
FND (IH, $\ell=\tau$)  &  1655    &    \\ 
\hline
\end{tabular}
\end{center}
\caption{  
The same as Table~3, but $\sqrt{s}=1$ TeV. 
} 
%\label{tab:2}
\end{table}
%%%%%%%%%%%%%%%%%%%%%%%%%%%%%%%%%%%%%%%%%%%%%%%%%%%%%%

In both $\sqrt{s}=500$ GeV and 1 TeV, 
 the signal to background ratio is large ($> 5-\sigma$) 
 for the FND (IH) and FD cases, 
 while the significance is negligible 
 for the FND (NH) case. 
If we naively expect a similar efficiency for the $\ell=\mu$ case, 
 the heavy neutrinos can be observed 
 with a large significance for both the FND and FD cases. 
In \cite{FujiiEtAl}, the $\ell=\tau$ case is also 
 analyzed in detail. 
In this case, the signal $N \to \nu e W (W \to q {\bar q}^\prime)$ 
 is considered as the background, and the analysis depends on 
 the number of the signal events and hence, 
 we cannot simply adopt the results in \cite{FujiiEtAl}. 
However, since the main backgrounds are 
 $e \nu W \to e \nu q \bar{q}$ and $WW \to \ell \nu q \bar{q}^\prime$ 
 also for this case, we can expect that 
 the efficiency for our case is similar to the one 
 obtained in \cite{FujiiEtAl}, 
 which is roughly the same as in the $\ell=e$ case. 
Thus, we expect, for the $\ell=\tau$ case, a large significance 
 for the signal events in both the FND and the FD cases.

%%%%%%%%%%%%%%%%%%%%%%%%%%%%%%%%%%%%%%%%%%%%%%%%%%%%%%%%%%%%%%%%%%%%%%%%
\subsection{Heavy neutrino signal with the general parameterization}
%%%%%%%%%%%%%%%%%%%%%%%%%%%%%%%%%%%%%%%%%%%%%%%%%%%%%%%%%%%%%%%%%%%%%%%%%%
In the general parameterization for the FND case, 
 $\cal{R}$ is a function of the Dirac phase ($\delta$), 
 the Majorana phase ($\rho$) and $y$ 
 in the general orthogonal matrix. 
In order to identify a region for these parameters 
 satisfying the constraint on the $\epsilon$-matrix, 
 we perform a parameter scan  
 by varying $-\pi \leq \delta, \rho \leq \pi$ 
 with an interval of $\frac{\pi}{20}$ and 
 $0 \leq y \leq 1$ with an interval of $0.02$.\footnote{
The Dirac mass matrix elements grow exponentially 
 as we raise $|y|$. 
For a value $y >1$, 
 the neutrino oscillation data are realized 
 under the fine-tuning between the large elements. 
Although the neutrino oscillation data are correctly 
 reproduced for any values of $y$ in the general parameterization, 
 we only consider $y \leq 1$ to avoid the fine-tuning. 
} 
Then, for the identified parameters, 
 we calculate the production cross section of 
 the $i$-th generation heavy neutrino at the LHC 
 through the process $q \bar{q}^\prime \to \ell N_i$  
 ($u \bar{d} \to \ell_\alpha^{+} N_i$ 
  and $ {\bar u} d \to \ell_\alpha^{-} \overline{N_i}$)
  given by 
\bea 
 \sigma(q \bar{q}^\prime \to \ell_\alpha N_i)
 = \sigma_{LHC} |{\cal R}_{\alpha i}(\delta,\rho,y)|^2,  
\eea
where $\sigma_{LHC}$ is the cross section given in Eq.~(\ref{XLHC}). 
Similarly, the production cross section at the ILC is 
\bea 
 \sigma(e^+ e^- \to \overline{\nu_\alpha} N_i)
 = \sigma_{ILC} |{\cal R}_{\alpha i}(\delta,\rho,y)|^2,  
\eea
where $\sigma_{ILC}$ is given in Eq.~(\ref{XILC}), 
 and we have used the approximation 
 ${\cal N}^\dagger {\cal R} \simeq U_{MNS}^\dagger {\cal R}$ 
 because $|\epsilon_{\alpha \beta}| \ll 1$ 
 as discussed in the previous section. 
The partial decay widths for the process 
 $N_i \to \ell_\alpha^{-} W^+/\nu_\alpha Z/ \nu_\alpha h$ 
 are obtained by multiplying Eq.~(\ref{widths}) 
 and the factor $|{\cal R}_{\alpha i}(\delta,\rho,y)|^2$ together.

\begin{figure}
\begin{center}
\includegraphics[scale=0.7]{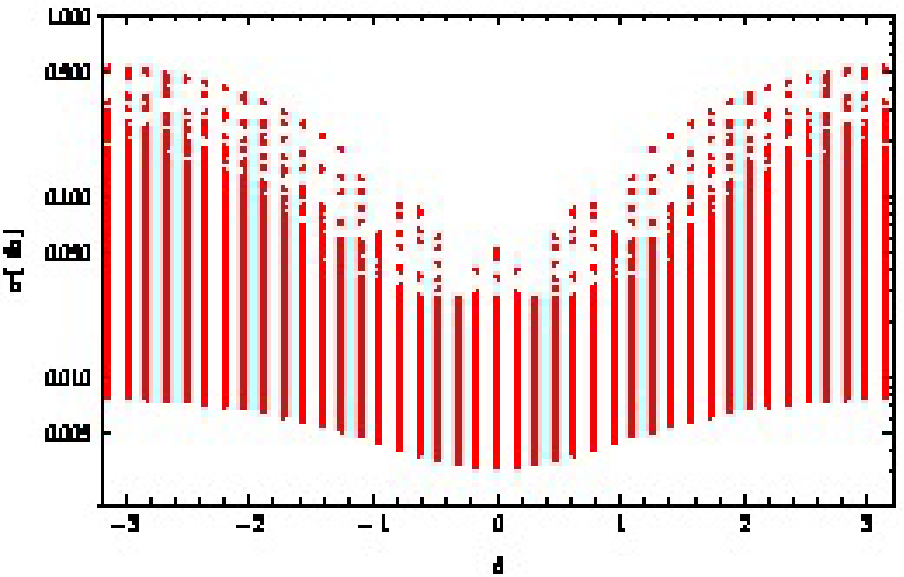}
\includegraphics[scale=0.7]{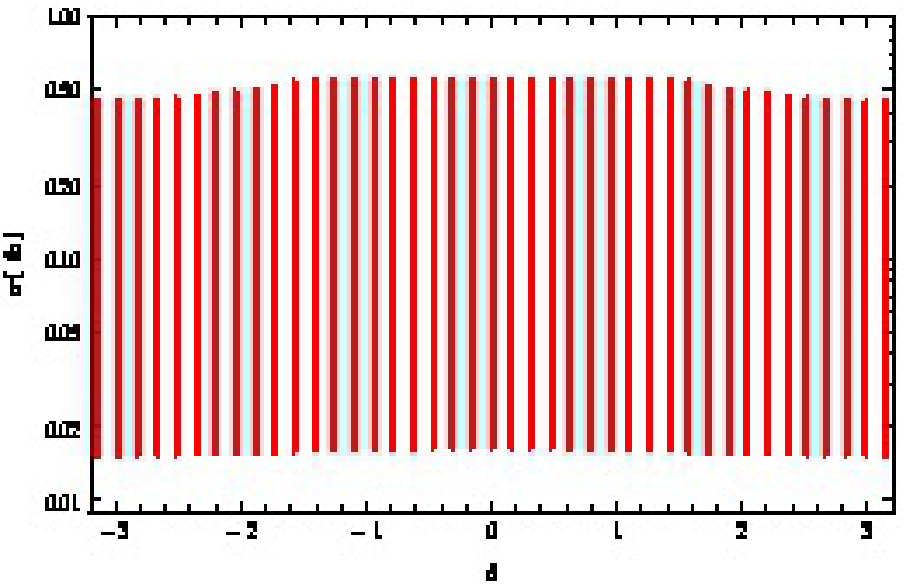}

\includegraphics[scale=0.7]{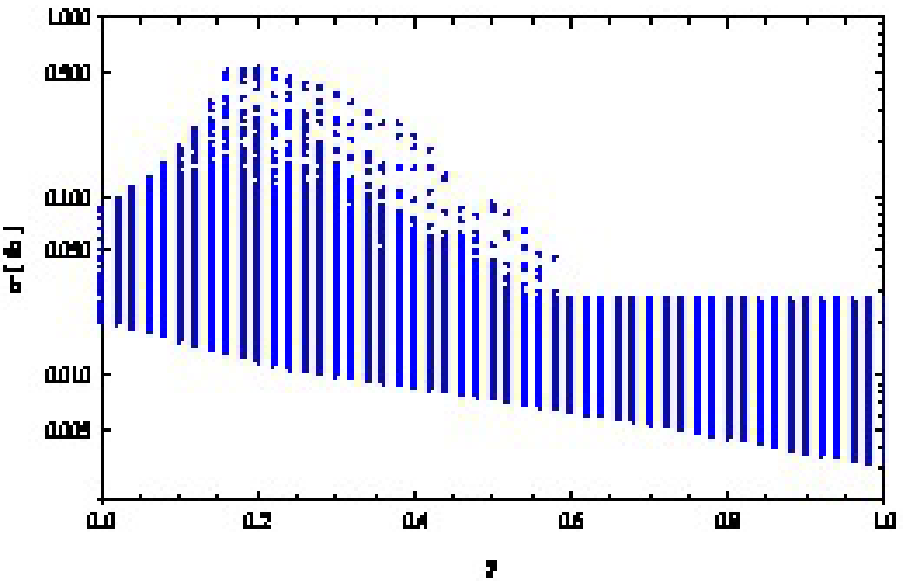}
\includegraphics[scale=0.7]{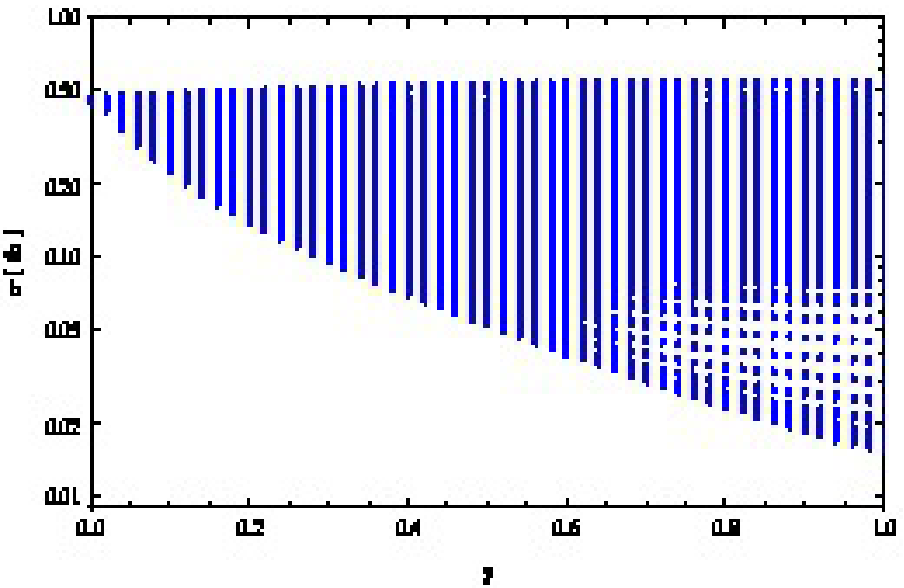}

\includegraphics[scale=0.7]{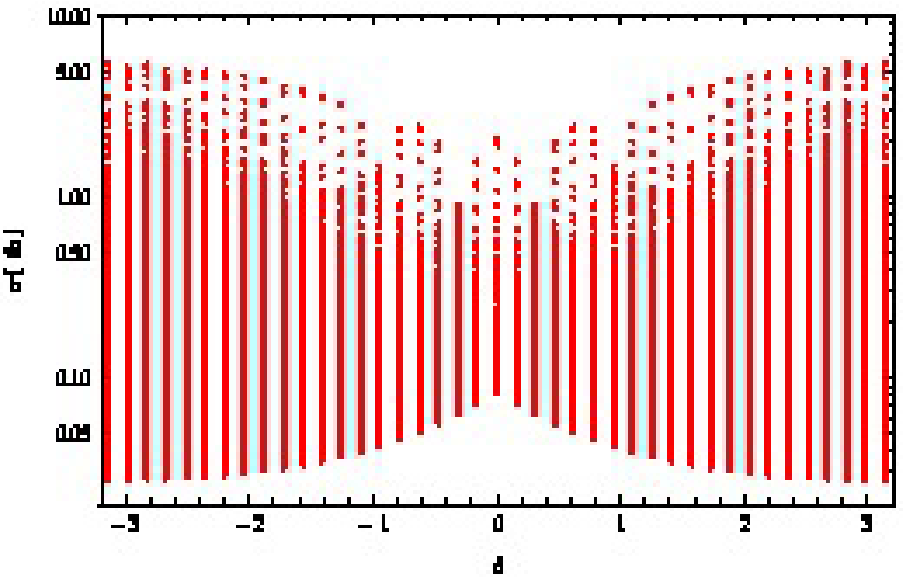}
\includegraphics[scale=0.7]{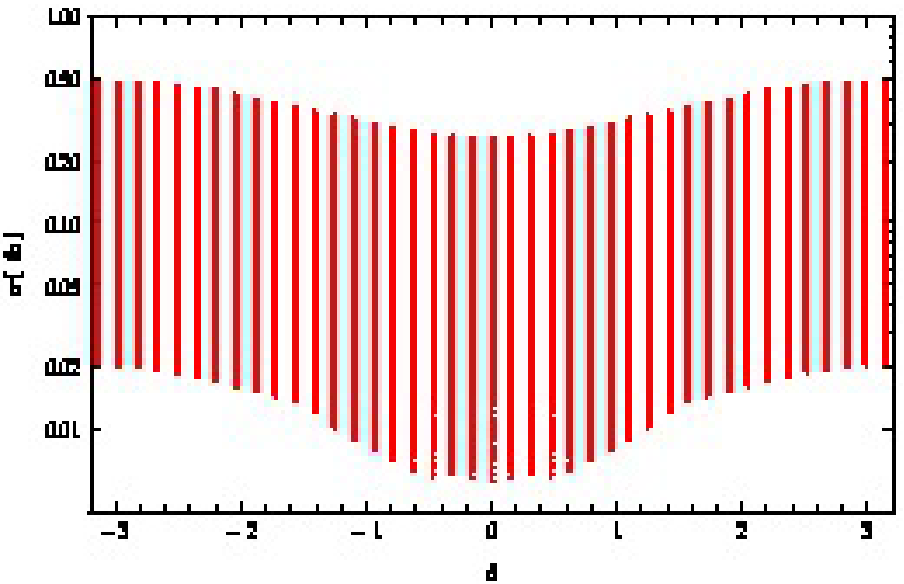}

\includegraphics[scale=0.7]{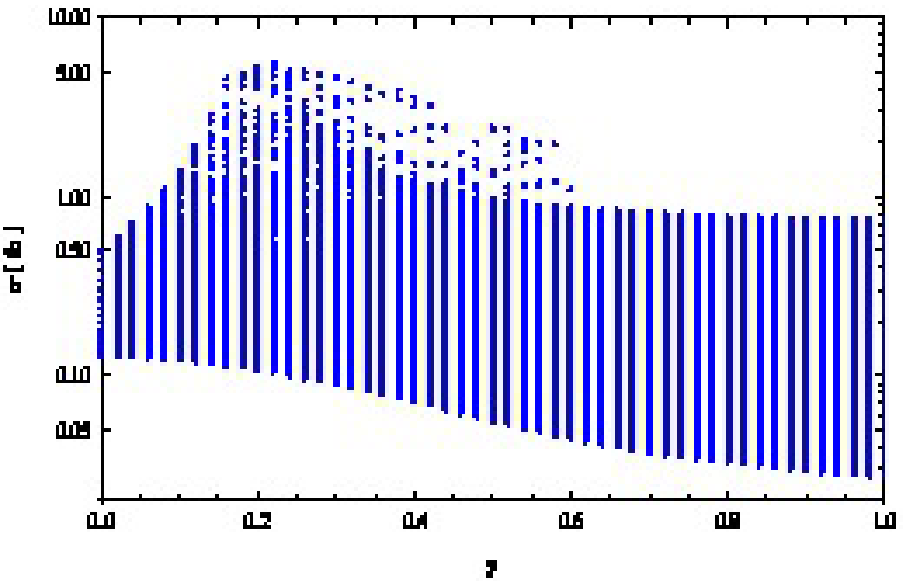}
\includegraphics[scale=0.7]{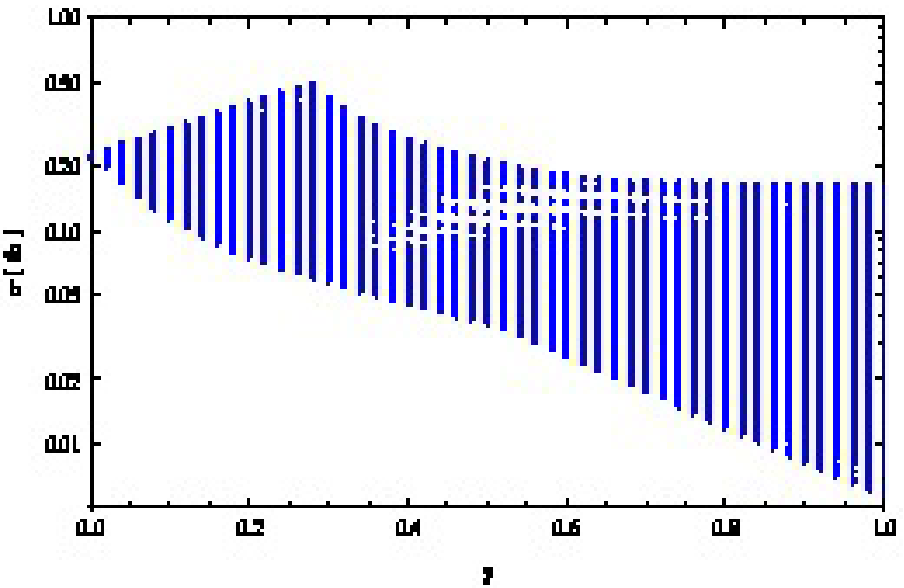}
\end{center}
\caption{
Signal cross sections providing the tri-lepton final states 
 as function of the Dirac phase ($\delta$) and $y$ 
 for the heavy neutrino mass of 100 GeV, 
 at the LHC with $\sqrt{s}=14$ TeV.
Each dot satisfies the experimental constraints 
 on all the elements in the $\epsilon$-matrix. 
The first (second) column corresponds to the results 
 for the NH (IH) case. 
The first two rows are for the final states with two electrons, 
 while the last tow are for the final states 
 with two muons. 
}
 \label{GenLHC}
\end{figure}

%%%%%%%%%%%%%%%%%%%%%%%%%%%%%%%%%%%%%%%%%%%%%%%%%%%%%%%%%%%%%%%%%%%%%%
%\subsubsection{Heavy neutrino signal
% for the general $\epsilon$ matrix  at LHC}
%%%%%%%%%%%%%%%%%%%%%%%%%%%%%%%%%%%%%%%%%%%%%%%%%%%%%%%%%%%%%%%%%%%%%%
Fig.~(\ref{GenLHC}) shows the results of the parameter scan 
 for the heavy neutrino production cross section 
 with the tri-lepton final states at the LHC. 
Each dots satisfies the experimental constraints 
 on all the $\epsilon$-matrix elements. 
The first (second) column shows the results 
 for the NH (IH) case. 
In the first (second) row, the results are shown 
 as a function of $\delta$ (y) for the final state  
 with two electrons, 
 while the corresponding results for the final state  
 with two muons are shown in the third and forth rows. 
Comparing the results with those for the simple parameterizations, 
 the signal cross sections for the NH case receive 
 significant enhancements for a certain parameter set, 
 while for the IH case, we only have an enhancement 
 by a factor $2-4$. 
The maximum signal cross sections we can achieve 
 in the general parameterization 
 are listed on Table~\ref{Max-LHC}. 
Interestingly, the maximum cross section for the NH case 
 with the final state including two muons 
 can even be larger than the one for the FD case.

%%%%%%%%%%%%%%%%%%%%%%%%%%%%%%%%%%%%
\begin{table}
\begin{center}
\begin{tabular}{c|cc}
\hline
         &  $ee$ & $\mu \mu$ \\ 
\hline 
      NH (fb) & $0.515$  & $5.95$  \\
      IH (fb) & $0.575$  & $0.475$  \\
\hline 
\end{tabular}
\end{center}
\caption{
The maximum LHC cross sections for the final states 
 with two electrons and two muons ,respectively, 
 at the LHC with $\sqrt{s}=14$ TeV.
} 
\label{Max-LHC}
\end{table}

%%%%%%%%%%%%%%%%%%%%%%%%%%%%%%%%%%%%%%%%%%%%%%%%%
%%%%%%%%%%%%%%%%%%%%%%%%%%%%%%%%%%%%%%%%%%%%%%%%%

%%%%%%%%%%%%%%%%%%%%%%%%%%%%%%%%%%%%%%%%%%%%%%%%%%%%%%%%%%
%\subsubsection{Heavy neutrino signal 
%for the general $\epsilon$ matrix  at ILC}
%%%%%%%%%%%%%%%%%%%%%%%%%%%%%%%%%%%%%%%%%%%%%%%%%%%%%%%%%%%

\begin{figure}
\begin{center}
\includegraphics[scale=0.7]{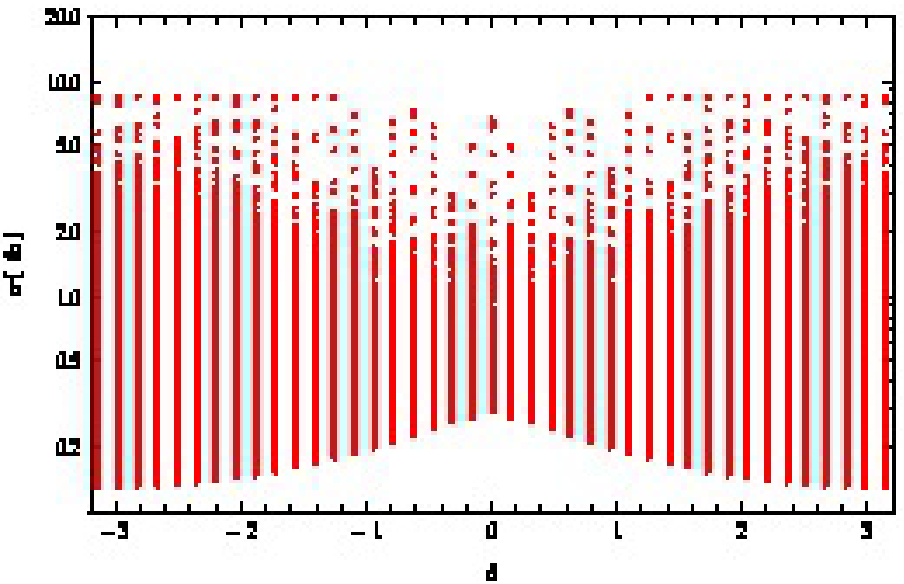}
\includegraphics[scale=0.7]{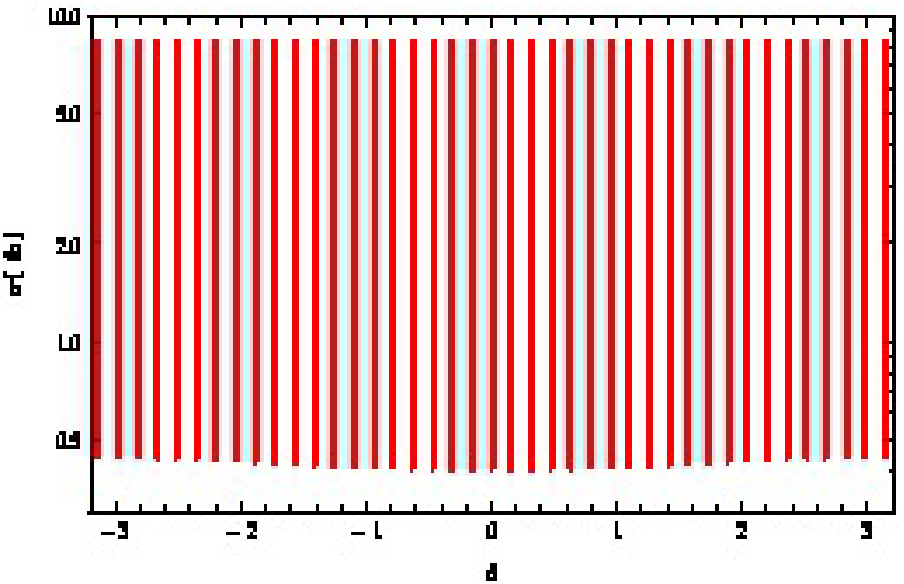}

\includegraphics[scale=0.7]{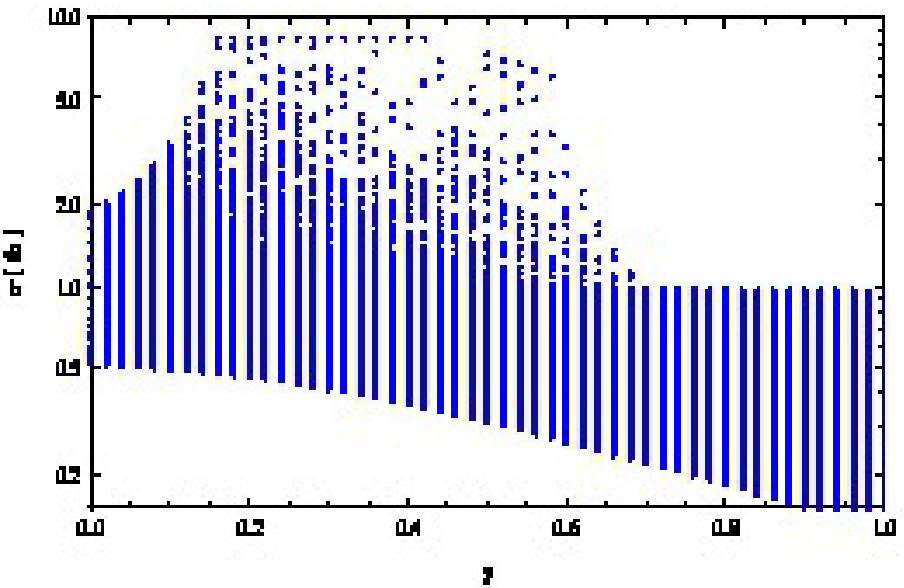}
\includegraphics[scale=0.7]{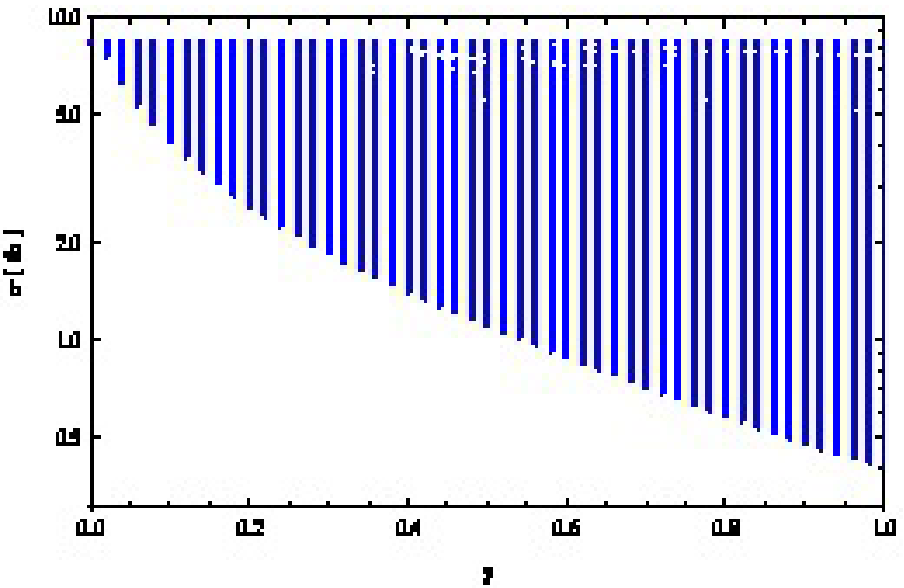}

\includegraphics[scale=0.7]{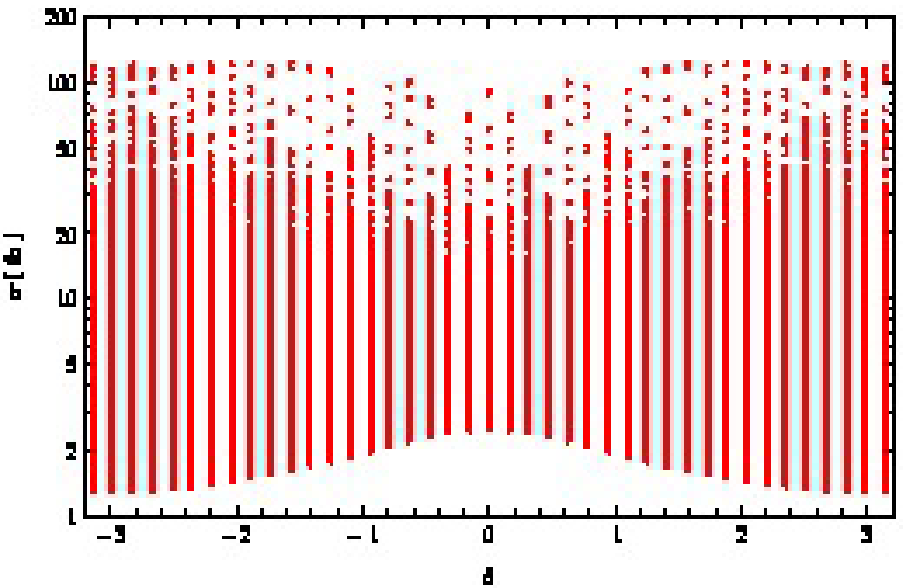}
\includegraphics[scale=0.7]{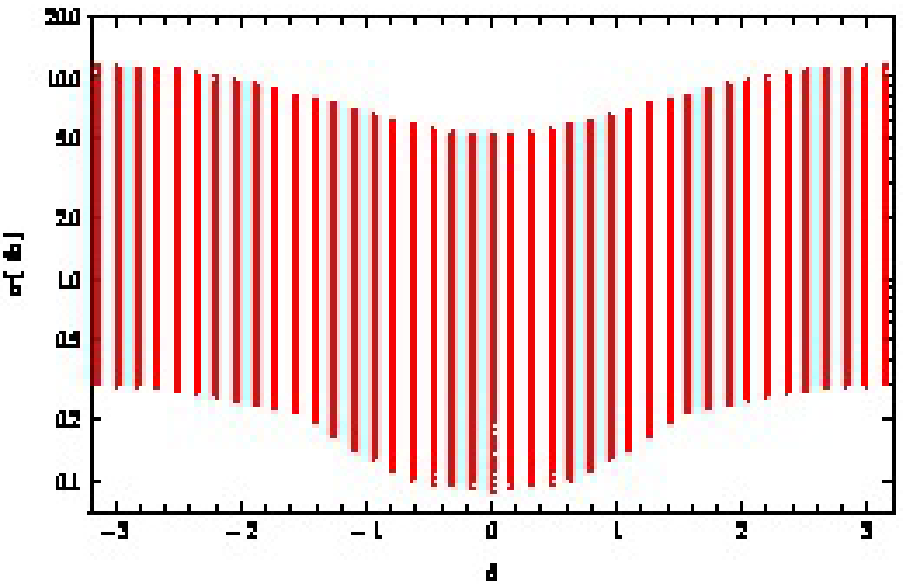}

\includegraphics[scale=0.7]{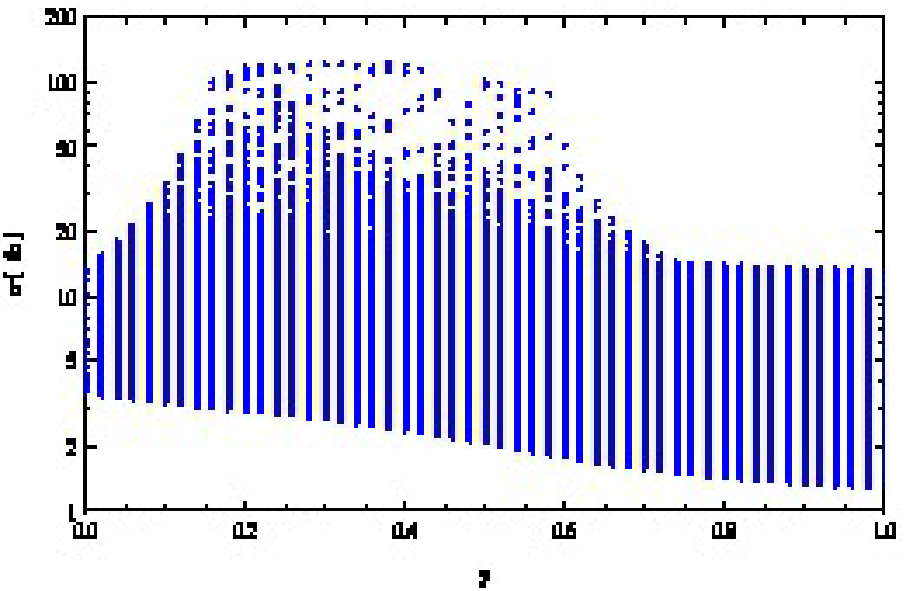}
\includegraphics[scale=0.7]{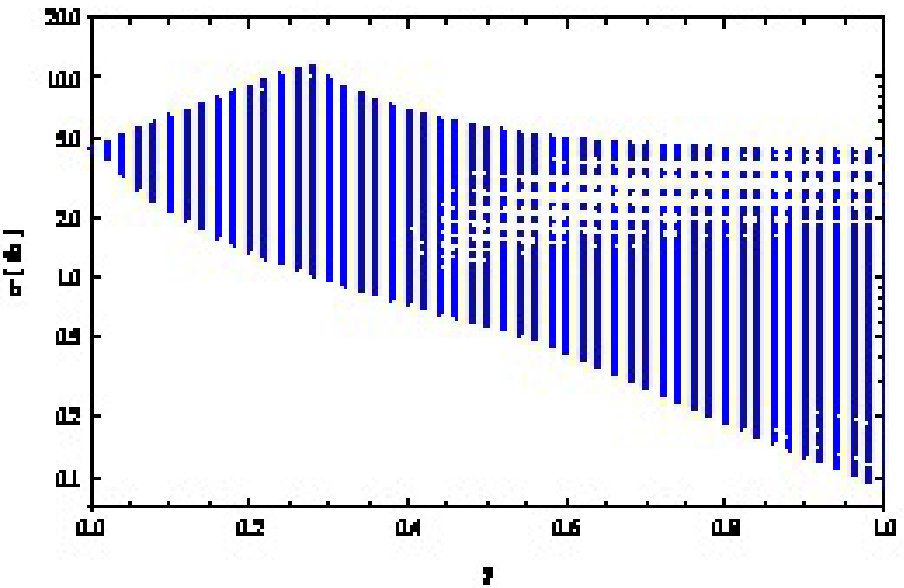}
\end{center}
\caption{
The production cross sections for the process 
 $e^+ e^- \to \nu N $, followed by the decays 
 $N \to \ell W$ ($\ell=e, \mu$) and $W \to q {\bar q}^\prime$, 
 as functions of the Dirac phase($\delta$) and $y$, 
 at the ILC with $\sqrt{s}=500$ GeV.  
Here we have fixed the heavy neutrino mass to be 150 GeV 
Each dots satisfies the experimental constraints 
 on all the $\epsilon$-matrix elements. 
The first (second) column shows the results 
 for the NH (IH) case. 
In the first (second) row, the results are shown 
 as a function of $\delta$ (y) for the case of $\ell=e$, 
 while the corresponding results for the case of $\ell= \mu$ 
 are shown in the third and forth rows. 
} 
\label{GenILC500}
\end{figure}

Fig.~\ref{GenILC500} shows the cross section 
 for the process $e^+e^- \to \nu N$, 
 followed by the decays $N \to \ell W$ and $W \to q {\bar q}^\prime$, 
 at the ILC with $\sqrt{s}=500$ GeV.  
Here we have fixed the heavy neutrino mass to be 150 GeV. 
Each dots satisfies the experimental constraints 
 on all the $\epsilon$-matrix elements. 
The first (second) column shows the results 
 for the NH (IH) case. 
In the first (second) row, the results are shown 
 as a function of $\delta$ (y) 
 for the case of $\ell=e$, 
 while the corresponding results for the case of $\ell= \mu$ 
 are shown in the third and forth rows. 
Similarly to the LHC results, 
 we have found the significant enhancements for the NH case 
 compared with the results for the simple parameterizations, 
 while we have no significant enhancement for the IH case. 
The maximum signal cross sections we can achieve 
 in the general parameterization 
 are listed on Table~\ref{Max-ILC500}. 
The maximum cross section for the NH case with $\ell=\mu$ 
 can even be larger than the one for the FD case. 
We have performed the same analysis also for 
 the ILC with $\sqrt{s}=1$ TeV. 
The maximum signal cross sections in this case 
 are listed on Table~\ref{Max-ILC1000}.  
We have about a $30-40$ \% enhancement 
 in the cross sections by the increase 
 of the collider energy. 

%%%%%%%%%%%%%%%%%%%%%%%%%%%%%%%%%%%%%%
\begin{table}
\begin{center}
\begin{tabular}{c|cc}
\hline
           &  NH (fb) & IH (fb)\\ 
\hline 
$\ell=e$   & $8.5$    & $8.5$    \\
$\ell=\mu$ & $130$    & $11.0$   \\ 
\hline 
\end{tabular}
\end{center}
\caption{
The maximum cross sections at the ILC with $\sqrt{s}=500$ GeV. 
Here we have fixed the heavy neutrino mass to be 150 GeV. 
Each dots satisfies the experimental constraints 
 on all the $\epsilon$-matrix elements. 
The first (second) column shows the results 
 for the NH (IH) case. 
The first and second rows correspond to 
 the results for the case of $\ell=e$, 
 while the corresponding results for the case of $\ell= \mu$ 
 are shown in the third and forth rows. 
} 
 \label{Max-ILC500}
\end{table}
%%%%%%%%%%%%%%%%%%%%%%%%%%%%%%%

%%%%%%%%%%%%%%%%%%%%%%%%%%%%%%%%%%%%%%
\begin{table}
\begin{center}
\begin{tabular}{c|cc}
\hline
           &  NH (fb) & IH (fb)\\ 
\hline 
$\ell=e$   & $11.0$    & $11.0$   \\
$\ell=\mu$ & $180$     & $180$    \\ 
\hline 
\end{tabular}
\end{center}
\caption{
The same as Table~\ref{Max-ILC500}, but for $\sqrt{s}=1$ TeV. 
} 
 \label{Max-ILC1000}
\end{table}
%%%%%%%%%%%%%%%%%%%%%%%%%%%%%%%

%%%%%%%%%%%%%%%%%%%%%%%%%%%%%%%%%%%%%%%%%%%%%%%%
%%%%%%%%%%%%%%%%%%%%%%%%%%%%%%%%%%%%%%%%%%%%%%%%%

%%%%%%%%%%%%%%%%%%%%%%%%%%%%%%%%%%%%%%%%%%%%%%%%%
%\begin{figure}
%\begin{center}
%\includegraphics[scale=0.7]{ILC_1000GeV_delta_EE_NH.eps}
%\includegraphics[scale=0.7]{ILC_1000GeV_delta_EE_IH.eps}
%
%\includegraphics[scale=0.7]{ILC_1000GeV_y_EE_NH.eps}
%\includegraphics[scale=0.7]{ILC_1000GeV_y_EE_IH.eps}
%
%\includegraphics[scale=0.7]{ILC_1000GeV_delta_MuMu_NH.eps}
%\includegraphics[scale=0.7]{ILC_1000GeV_delta_MuMu_IH.eps}
%
%\includegraphics[scale=0.7]{ILC_1000GeV_y_MuMu_NH.eps}
%\includegraphics[scale=0.7]{ILC_1000GeV_y_MuMu_IH.eps}
%\end{center}
%\caption{The production cross sections for the process 
% $e^+ e^- \to \nu N $, followed by the decays 
% $N \to \ell W$ ($\ell=e, \mu, \tau$) and $W \to q {\bar q}^\prime$, 
% as functions of the Dirac phase($\delta$) and y keeping the heavy neutrino mass fixed at 150 GeV with $\sqrt{s}=1000$ GeV. 
%The first and second row show the e+jj final states for the NH and IH cases respectively. The third and fourth rows show the $\mu$+jj final states for the NH and IH cases respectively..}
%\label{GenILC2}
%\end{figure}
%

The main backgrounds are 
 $e \nu W \to e \nu q \bar{q}$ and $WW \to \ell \nu q \bar{q}$, 
 which are dramatically reduced by the selection 
 using an isolated-electron track with 
 a requirement of its energy range, 10 GeV$\leq E_e \leq 200$ GeV, 
 the requirement of the reconstructed  di-jet mass to be 
 consistent with W hypothesis etc (see \cite{FujiiEtAl} for details).
The Maximum signal cross section
 for $\sqrt{s}=1$ TeV and the same luminosity ${\cal L}=500$ fb$^{-1}$  
 are listed on Table~7. We have only listed $e$+jj and  $\mu$+jj signal cross section as functions of $\delta$, $\rho$ and $y$.

From Tables~6 and 7 the signal cross sections for $\mu$+jj in NH dominates over IH by an order of magnitude for both the collider energies, $\sqrt{s}=500$GeV and $\sqrt{s}=1$ TeV. The signal cross sections for e+jj in NH is almost the same as that in the IH case for both the collider energies, $\sqrt{s}=500$GeV and $\sqrt{s}=1$ TeV. The $\mu\mu$ case cross sections at 
$\sqrt{s}=500$GeV and $\sqrt{s}=1$ TeV are some factors greater than the FD cases respectively. 
%%%%%%%%%%%%%%%%%%%%%%%%%%%%%%%%%%%%%%%%%%%%%%%%%

%%%%%%%%%%%%%%%%%%%%%%%%%%%%%%%%%%%%%%%%%%
\section{Conclusions}
%%%%%%%%%%%%%%%%%%%%%%%%%%%%%%%%%%%%%%%%%%
We have studied the inverse seesaw scenario 
 and the signature of the pseudo-Dirac heavy neutrino 
 production at the LHC and ILC. 
In the inverse seesaw scenario, the light neutrino masses 
 are realized by small lepton-number-violating parameters 
 and hence the SM singlet neutrinos have sizable 
 Dirac Yukawa couplings with the SM lepton doublets and 
 Higgs doublet even for their mass scale being at the TeV scale or smaller. 
As a result, the heavy  neutrinos can be produced at the LHC and ILC. 
Based on a concrete model realizing the inverse seesaw 
 in the context of the NMSSM, we have fixed the model parameters 
 so as to satisfy the experimental results 
 such as the neutrino oscillation data, 
 the precision measurements of the weak boson decays, 
 and the lepton-flavor-violating decays of charged leptons. 
We have considered two typical cases for the neutrino flavor structures 
 of the model, namely, the FND and FD cases. 
With the fixed parameters, we have calculated the production 
 cross sections of the heavy neutrinos at the LHC and ILC.

First we have considered simple parameterizations 
 with all zero CP-phases . 
For the LHC with $\sqrt{s}=14$ TeV, 
 we have analyzed the productions of the heavy neutrinos 
 with a degenerate 100 GeV mass, providing 
 the tri-lepton final states with the like-sign electrons or muons. 
After imposing suitable cuts, 
 we have found that the $5-\sigma$ statistical significance 
 of the signal events over the SM background can be achieved 
 for the luminosity around 11 fb$^{-1}$ in the FD case. 
On the other hand, the production cross sections in the FND case 
 is too small to observe the heavy neutrino signal.

We have also studied the heavy neutrino production 
 at the ILC with $\sqrt{s}=$500 GeV-1 TeV, 
 where the final state with a single, isolated electron, 
 and di-jet and large missing energy is considered. 
For the luminosity $\sqrt{s}=500$ fb$^{-1}$, 
 we can obtain clear signatures of the heavy neutrinos 
 with mass 150 GeV for the IH mass spectrum 
 in the FND case and the FD case. 
On the other hand, the significance 
 for the NH mass spectrum in the FND case has been found to be low. 
Since we can expect the similar efficiencies of the signal 
 and SM background for the final states with different 
 lepton flavors, muon or tau, 
 the heavy neutrinos can be detected with a large statistical
 significance in the modes for all FND and FD cases.

For completeness, we have also considered the general parameterization 
 for the Dirac neutrino mass matrix by introducing 
 a general orthogonal matrix and CP-phases, for the FND case. 
In this case, three new parameters, the Dirac CP-phase ($\delta$),  
 the Majorana CP-phase ($\rho$) and one angle of the orthogonal matrix, 
 are newly involved in our analysis. 
We have performed a parameter scan and identified 
 the parameter region which satisfies all experimental 
 constraints on the elements of the $\epsilon$-matrix. 
Then, we have shown the signal cross sections 
 of the heavy neutrino production for the parameters identified. 
For both the LHC and ILC cases, we have found significant
 enhancements of the cross section for the NH case  
 and the resultant cross section can be of the same order  
 of the FD case. 
On the other hand, such a remarkable enhancement 
 has not been observed for the IH case.

If the heavy neutrinos are discovered in the future, 
 this indicates that a mechanism of the neutrino mass generation 
 is not due to the conventional seesaw mechanism, 
 because the expected cross section for the conventional  
 seesaw is extremely small. 
In addition, the flavor dependent signal events 
 from the heavy neutrino productions 
 provide us with valuable information 
 to investigate the flavor structure of the model 
 for the neutrino mass generation.

Finally we comment on the current bound of the heavy neutrino 
 production at the LHC. 
The ATLAS experiment~\cite{ATLAS-N} has reported their results 
 on the search for the heavy neutrinos based on the  
 production through effective four-fermion operators~\cite{4Fermi}. 
The vector operator of 
 $ ({\bar d} \gamma^\mu u)({\bar N} \gamma_\mu \ell)/\Lambda^2 $ 
 is relevant to our case. 
The final states with $\ell \ell jj$ ($\ell=e$ or $\mu$) 
 have been analyzed as a signal of the heavy neutrino production, 
 followed by the decay $N \to \ell W$, $W \to jj$. 
 From the data corresponding to an integrated luminosity 
 of 2.1 fb$^{-1}$ at $\sqrt{s}=7$ TeV, 
 the ATLAS experiment has set the lower bound on 
 the cutoff scale $\Lambda$ as a function of the heavy neutrino 
 mass $\geq 200$ GeV. 
For example, it is found that $\Lambda \geq 2.8$ TeV for $M=200$ GeV. 
We interpret this result to the upper bound on the heavy neutrino 
 production cross section through the four-fermion operator 
 as $\sigma(q {\bar q}^\prime \to \ell N) \leq 24.0$ fb. 
In the FD case, we find 
 $\sigma(q {\bar q}^\prime \to \ell N) \simeq 3.77$ fb 
 and therefore, the parameter region we have examined 
 in this paper is consistent with the current LHC results.

%%%%%%%%%%%%%%%%%%%%%%%%%%%%%%%%%%
\section*{Acknowledgments}
%%%%%%%%%%%%%%%%%%%%%%%%%%%%%%%%%%
A.D. would like to thank Joydeep Chakrabortty, 
 Partha Konar and Srubabati Goswami 
 for useful discussions and comments. 
He would also like to thank PRL, Ahmedabad 
 for their hospitality when a part of the work is done.
This visit is supported by APS-IUSSTF Physics Student 
 Visitation Program. 
The work of N.O. is supported in part 
 by the DOE Grants, No. DE-FG02-10ER41714.
%%%%%%%%%%%%%%%%%%%%%%%%%%%%%%%%%%

%%%%%%%%%%%%%%%%%%%%%%%%%%%%%%%%%%%


\begin{thebibliography}{99}
%%%%%%%%%%%%%%%%%%%%%%%%%%%%%%%%%%%

\bibitem{PDG}
K.~Nakamura {\it et al.}  [Particle Data Group Collaboration],
  %``Review of particle physics,''  
 J.\ Phys.\ G G {\bf 37}, 075021 (2010).


\bibitem{T2K}
K. Abe et. al. [T2K Collaboration] 
  Phys.\ Rev.\ Lett.\ { \bf 107}, 041801 (2011). 
%  [arXiv:1106.2822[hep-ex]]


\bibitem{MINOS}
P.~Adamson {\it et al.}  [MINOS Collaboration],
  %``Improved search for muon-neutrino to electron-neutrino oscillations in MINOS,''
  Phys.\ Rev.\ Lett.\  {\bf 107}, 181802 (2011).
%  [arXiv:1108.0015 [hep-ex]].


\bibitem{DCHOOZ} 
Y.~Abe {\it et al.}  [DOUBLE-CHOOZ Collaboration],
  %``Indication for the disappearance of reactor electron antineutrinos in the Double Chooz experiment,''
  Phys.\ Rev.\ Lett.\  {\bf 108}, 131801 (2012).
%   [arXiv:1112.6353 [hep-ex]].


\bibitem{DayaBay}
F.~P.~An {\it et al.}  [DAYA-BAY Collaboration],
  %``Observation of electron-antineutrino disappearance at Daya Bay,''
  Phys.\ Rev.\ Lett.\  {\bf 108}, 171803 (2012).
%  [arXiv:1203.1669 [hep-ex]].


\bibitem{RENO}
J.~K.~Ahn {\it et al.}  [RENO Collaboration],
  %``Observation of Reactor Electron Antineutrino Disappearance in the RENO Experiment,''
  Phys.\ Rev.\ Lett.\  {\bf 108}, 191802 (2012). 
%   [arXiv:1204.0626 [hep-ex]].


\bibitem{Seesaw}
P.~Minkowski, Phys. Lett. B {\bf 67}, 421 (1977);
T.~Yanagida, in \emph{Proceedings of the Workshop on the Unified
  Theory and the Baryon Number in the Universe} (O.~Sawada and
  A.~Sugamoto, eds.), KEK, Tsukuba, Japan, 1979, p.~95;
M.~Gell-Mann, P.~Ramond, and R.~Slansky, \emph{Supergravity} (P.~van
  Nieuwenhuizen et al. eds.), North Holland, Amsterdam, 1979, p.~315;
S.~L. Glashow, \emph{The future of elementary particle physics}, in
  \emph{Proceedings of the 1979 Carg{\`e}se Summer Institute
 on Quarks and Leptons} (M.~Levy et al. eds.),
 Plenum Press, New York, 1980, p.~687;
R.~N. Mohapatra and G.~Senjanovic,
 Phys. Rev. Lett. {\bf 44}, 912 (1980).


\bibitem{InvSeesaw}
R.~N.~Mohapatra,
  %``MECHANISM FOR UNDERSTANDING SMALL NEUTRINO MASS IN SUPERSTRING THEORIES,''
  Phys.\ Rev.\ Lett.\  {\bf 56} (1986) 561;
%
R.~N.~Mohapatra and J.~W.~F.~Valle,
  %``NEUTRINO MASS AND BARYON-NUMBER NONCONSERVATION IN SUPERSTRING MODELS,''
  Phys.\ Rev.\  D {\bf 34}, 1642 (1986).


\bibitem{GOS} 
  I.~Gogoladze, N.~Okada and Q.~Shafi,
  %``NMSSM and Seesaw Physics at LHC,''
  Phys.\ Lett.\ B {\bf 672}, 235 (2009).
%   [arXiv:0809.0703 [hep-ph]].


\bibitem{NMSSM}
 P.~Fayet, Nucl.\ Phys.\ {\bf B 90} (1975) 104;
 H.~P.~Nilles, M.~Srednicki and D.~Wyler,
   Phys.\ Lett.\ {\bf B 120} (1983) 346;
 J.~P.~Derendinger and C.~A.~Savoy, Nucl.\ Phys.\ {\bf B 237} (1984) 307;
 J.~ Ellis, J.~F.~Gunion, H.~E.~Haber, L.~Roszkowski and F.~Zwirner,
 Phys.\ Rev.\ {\bf D 39} (1989) 844;
 L.~Durand and J.~L.~Lopez, Phys.\ Lett.\ {\bf B 217} (1989) 463;
 M.~Drees, Int.\ J.\ Mod.\ Phys.\ {\bf A 4} (1989) 3635.


\bibitem{dim6I}
  A.~Broncano, M.~B.~Gavela and E.~E.~Jenkins,
  %``The effective Lagrangian for the seesaw model of neutrino mass and
  %leptogenesis,''
  Phys.\ Lett.\  B {\bf 552}, 177 (2003)
  [Erratum-ibid.\  B {\bf 636}, 330 (2006)];
 %   [arXiv:hep-ph/0210271];
%  A.~Broncano, M.~B.~Gavela and E.~E.~Jenkins,
  %``Neutrino physics in the seesaw model,''
  Nucl.\ Phys.\  B {\bf 672}, 163 (2003).
 %   [arXiv:hep-ph/0307058].


\bibitem{CTEQ}
J.~Pumplin, D.~R.~Stump, J.~Huston, H.~L.~Lai, P.~Nadolsky and W.~K.~Tung,
  %``New generation of parton distributions with uncertainties from global  QCD
  %analysis,''
  JHEP {\bf 07} (2002) 012. 


\bibitem{ATLAS}
 G.~Aad et al. [ATLAS Collaboration], Phys. Lett. B {\bf716}, 1 (2012)


\bibitem{CMS}
 S.~Chatrchyan et al. [CMS Collaboration], Phys. Lett. B {\bf716}, 30 (2012)


\bibitem{Constraints1}
  S.~Antusch, C.~Biggio, E.~Fernandez-Martinez, M.~B.~Gavela and J.~Lopez-Pavon,
  %``Unitarity of the Leptonic Mixing Matrix,''
  JHEP {\bf 0610}, 084 (2006).
 %  [arXiv:hep-ph/0607020].


\bibitem{Constraints2}
  A.~Abada, C.~Biggio, F.~Bonnet, M.~B.~Gavela and T.~Hambye,
  %``Low energy effects of neutrino masses,''
  JHEP {\bf 0712}, 061 (2007).
  %  [arXiv:0707.4058 [hep-ph]].


\bibitem{Constraints3}
A.~Ibarra, E.~Molinaro and S.~T.~Petcov,
  %``TeV Scale See-Saw Mechanisms of Neutrino Mass Generation, the Majorana Nature of the Heavy Singlet Neutrinos and $(\beta\beta)_{0\nu}$-Decay,''
  JHEP {\bf 1009}, 108 (2010); 
%   [arXiv:1007.2378 [hep-ph]].
%
%  A.~Ibarra, E.~Molinaro and S.~T.~Petcov,
%  %``Low Energy Signatures of the TeV Scale See-Saw Mechanism,''
  Phys.\ Rev.\ D {\bf 84}, 013005 (2011); 
%  [arXiv:1103.6217 [hep-ph]].
%
D.~N.~Dinh, A.~Ibarra, E.~Molinaro and S.~T.~Petcov,
  %``The $\mu - e$ Conversion in Nuclei, $\mu \to e \gamma, \mu \to 3e$ Decays and TeV Scale See-Saw Scenarios of Neutrino Mass Generation,''
  JHEP {\bf 1208}, 125 (2012)
  [Erratum-ibid.\  {\bf 1309}, 023 (2013)].
%  [arXiv:1205.4671 [hep-ph]].


\bibitem{Adam}
J.~Adam et. al. [MEG Collaboration], 
Phys.\ Rev.\ Lett. {\bf 107},171801, (2011). 
%   [arXiv: 1107.5541 [hep-ex]]

\bibitem{Aubert}
B.~Aubert et. al. [BABAR Collaboration],  
Phys.\ Rev.\ Lett. {\bf 104},021802,(2010). 
%    [arXiv: 0908.2381[hep-ex]]

\bibitem{OLeary}
See, for summary, 
 B.~O' Leary et. al. [SuperB Collaboration], 
 arXiv: 1008.1541[hep-ex]. 



\bibitem{tri-lepton}
F.~del Aguila and J.~A.~Aguilar-Saavedra,
  %``Distinguishing seesaw models at LHC with multi-lepton signals,''
  Nucl.\ Phys.\ B {\bf 813}, 22 (2009); 
%  [arXiv:0808.2468 [hep-ph]].
%
%F.~del Aguila and J.~A.~Aguilar-Saavedra,
  %``Electroweak scale seesaw and heavy Dirac neutrino signals at LHC,''
  Phys.\ Lett.\ B {\bf 672}, 158 (2009). 
%   [arXiv:0809.2096 [hep-ph]].





\bibitem{mue1}
M.~Hirsch, F.~Staub and A.~Vicente,
  %``Enhancing $l_i \to 3 l_j$ with the $Z^0$-penguin,''
  Phys.\ Rev.\ D {\bf 85}, 113013 (2012).
%   [arXiv:1202.1825 [hep-ph]].


\bibitem{mue2}
A.~Abada, D.~Das, A.~Vicente and C.~Weiland,
  %``Enhancing lepton flavour violation in the supersymmetric inverse seesaw beyond the dipole contribution,''
 JHEP {\bf09} (2012)015. 
%arXiv:1206.6497 [hep-ph].



\bibitem{tri-lepton2}
C.~Y.~Chen and P.~S.~B.~Dev,
  %``Multi-Lepton Collider Signatures of Heavy Dirac and Majorana Neutrinos,''
  Phys.\ Rev.\ D {\bf 85}, 093018 (2012). 
%   [arXiv:1112.6419 [hep-ph]].


\bibitem{FujiiEtAl}
T.~Saito, M.~Asano, K.~Fujii, N.~Haba, S.~Matsumoto, T.~Nabeshima, Y.~Takubo and H.~Yamamoto {\it et al.},
  %``Extra dimensions and Seesaw Neutrinos at the International Linear Collider,''
  Phys.\ Rev.\ D {\bf 82}, 093004 (2010).
%  [arXiv:1008.2257 [hep-ph]].


\bibitem{XdimNu}
N.~Haba, S.~Matsumoto and K.~Yoshioka,
  %``Observable Seesaw and its Collider Signatures,''
  Phys.\ Lett.\ B {\bf 677}, 291 (2009).
%  [arXiv:0901.4596 [hep-ph]].


\bibitem{ATLAS-N}
G.~Aad {\it et al.}  [ATLAS Collaboration],
  %``Search for heavy neutrinos and right-handed W bosons in events with two leptons and jets in pp collisions at sqrt(s) = 7 TeV with the ATLAS detector,''
  Eur.\ Phys.\ J.\ C {\bf 72}, 2056 (2012).
%   [arXiv:1203.5420 [hep-ex]].


\bibitem{4Fermi}
F.~del Aguila, S.~Bar-Shalom, A.~Soni and J.~Wudka,
  %``Heavy Majorana Neutrinos in the Effective Lagrangian Description: Application to Hadron Colliders,''
  Phys.\ Lett.\ B {\bf 670} (2009) 399. 
%  [arXiv:0806.0876 [hep-ph]].


\end{thebibliography}
\end{document}